# Fluorophore signal detection and imaging enhancement in high refractive index nanowire biosensors

Nicklas Anttu[1]

[1]Physics, Faculty of Science and Engineering, Åbo Akademi University, FI-20500 Turku, Finland

Email: nicklas.anttu@abo.fi

## Abstract

High refractive index semiconductor nanowires have recently been demonstrated experimentally as an efficient platform for enhancing the signal in fluorescence-based biosensors. Here, we study through modelling how a vertical GaP nanowire (i) enhances the excitation intensity at the position of the fluorophore attached to the nanowire sidewall, (ii) enhances the probability to collect photons emitted from the fluorophore by directing them preferentially into the numerical aperture of the collection objective, and (iii) through the Purcell effect increases the quantum yield of the fluorophore. With appropriate choice for the geometry of the nanowire, we can reach a larger than $10^2$ enhancement in signal compared to a corresponding conventional planar biosensor platform. We model also imaging-based detection. There, we find that thanks to waveguiding in the nanowire, we can beat the limitations set by the depth of view in conventional microscopy, enabling the use of a long nanowire to enhance the binding-area for fluorophores. As an example, we can focus to the top of a 4000 nm long nanowire and reach a 25 times sharper image from a fluorophore at the bottom of the nanowire, as compared to such a 4000 nm defocusing in a conventional planar biosensor platform.

## Introduction

Biosensors based on fluorescent labels are used for reliable and reproducible detection of low-concentrations of biomarkers with high specificity.[1,2] In a typical configuration, such a fluorescence-based biosensors is fabricated within a flow cell to allow flushing of liquids through the biosensor for (i) functionalization of the surface and (ii) consecutive binding of biomarkers from biological fluid-samples to the biosensor surface, (iii) followed by flushing of fluorophores that bind selectively to the biomarkers. Fluorescence-based biosensors find currently use, for example, in cancer,[3] asthma,[4] and heart disease diagnostics.[5] An even more sensitive biosensor would be beneficial for example for detection of biomarkers at low concentration at early stages of diseases.

Elongated nanostructures, in the form of high-refractive index semiconductor nanowires aligned vertical to the optical axis (see Figure 1 for a schematic), offer optical effects beyond those available in conventional, planar systems. Vertical semiconductor nanowires can be fabricated epitaxially[6,7] or through top-down etching,[8] giving large materials and geometry freedom. Semiconductor nanowires are popular for optical and optoelectronic applications like solar cells,[9–11] photodetectors,[12,13] LEDs,[14,15] and lasers.[16–18]

For semiconductor nanowire based fluorescent biosensors, various materials, including GaP,[1] GaAs,[19] and ZnO[20–22] have been employed. A very promising enhancement in the signal by a factor of 10 was recently demonstrated in experiments with GaP nanowire-based fluorescent biosensors, as compared to the industry-standard planar biosensor.[1] Thanks to diffraction effects enabled by the high refractive index, semiconductor nanowires can enhance the excitation and emission properties of fluorophores residing in the vicinity of the nanowire surface (see Figure 1 for schematic).[1,19,20,23–28] Such modifications are sensitive to the choice for the nanowire diameter.[19,20,25,28] For imaging-based detection, nanowires with appropriately chosen diameter offer lightguiding properties, which allow to focus toward the tip of the nanowire to capture a sharp image of the fluorescence, irrespective of the actual binding position of the fluorophore along the nanowire sidewall.[24,25,28]

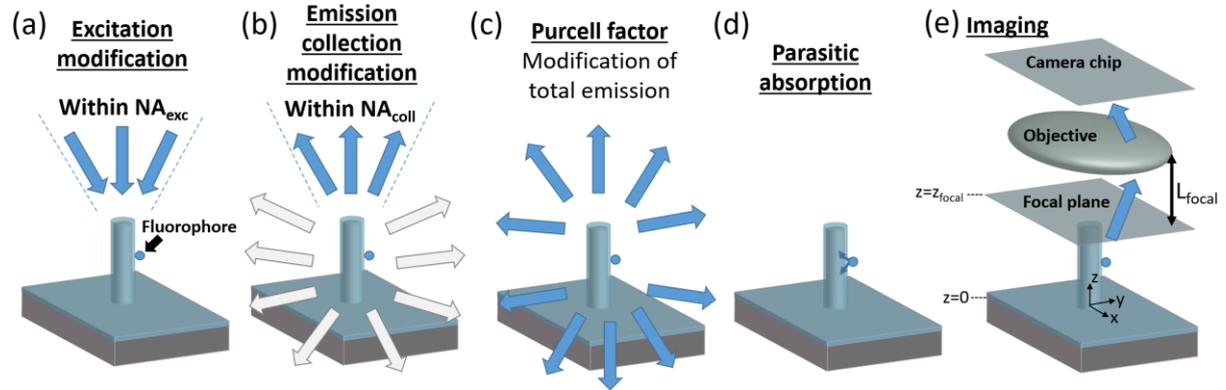

**Figure 1**. (a)-(d) Schematic of the four optical effects and (e) image formation that we model and analyse for fluorescence-based nanowire biosensors. In (e), the camera chip indicates the image plane.

A nanowire-based system has too many design parameters to allow efficient optimization for biosensing applications through prototyping and characterization in the lab, for example the choice of nanowire material, diameter $D$ and length $L$ of nanowire, excitation wavelength $\lambda_{exc}$, fluorophore emission wavelength $\lambda_{em}$, and numerical aperture $NA_{exc}$ for excitation light and $NA_{coll}$ for fluorophore emission. Optics modelling is an efficient tool for analysing the optical response of such nanowire-platforms.[19,20,25–30] However, modelling of the optical response of a nanowire-based fluorescent biosensor for varying geometry, materials, $\lambda_{exc}$, $\lambda_{em}$, $NA_{exc}$, and $NA_{coll}$ has not been rigorously addressed, especially including imaging-based detection.

Here, we develop and employ such modelling for design and analysis of nanowire-based fluorescent biosensors (the five optical effects depicted in Figure 1). With appropriate choice for the geometry of the nanowire, in combination with the choice for the excitation and emission conditions, we can reach a larger than $10^2$ enhancement in signal compared to a corresponding conventional planar biosensor platform. We expect to reach a high signal detection level even with a small NA for the combined excitation and detection system. Thus, the nanowire-based biosensor offers the prospect of high-signal biosensing with a portable, low-cost low-NA optical read-out system, even out in the field, i.e., a lab-on-a-chip system.[1] Furthermore, fluorescence-based biodetectors are typically used in imaging mode, and we develop also such modelling. There, we find that thanks to waveguiding in the nanowire, we can beat the limitations set by the depth of view in conventional microscopy, enabling the use of a long nanowire to enhance the binding-area for fluorophores. As an example, we can focus to the top of a 4000 nm long nanowire and reach a 25 times sharper image from a fluorophore at the bottom of the nanowire, as compared to such a 4000 nm defocusing in a conventional planar biosensor platform.

## Geometry and optics model

We consider light in a wavelength range of 400 < $\lambda$ < 900 nm for the excitation light and fluorescence emission, at wavelength $\lambda_{exc}$ and $\lambda_{em}$, respectively. This wavelength range covers a broad range of fluorophores of interest.[31] The light-scattering is described by the Maxwell equations, and we take into account the optical response of the constituent materials through their refractive index $n$.[32] We solve numerically for the light-scattering in the three-dimensional nanowire system (see Supporting Information for details).

## Geometry and refractive indexes

Unless otherwise stated, we study a GaP nanowire on top of a GaP substrate, with a liquid of refractive index $n$ = 1.33, that is, similar to that of water, surrounding the nanowire (see Figure 1 for a schematic). The length of the nanowire is $L$ and the diameter is $D$. On top of both the nanowire and the substrate, we have a thin, 10 nm thick, SiO$_2$ coating to ascertain biocompatibility of the surface.[1] For the SiO$_2$, we use tabulated values[33] of $n \approx 1.45$, and for GaP tabulated values from Ref. [34].

For GaP, Re($n$) ≈ 3.4 in the wavelength range 400 < λ < 900 nm that we study (see Figure S3 in the Supporting Information). On the other hand, Im($n$) of the GaP gives rise to absorption, which can be quantified for example through the absorption length $L_{abs}$ of bulk GaP. For GaP, the band-to-band absorption seizes at the bandgap wavelength of ≈550 nm beyond which GaP is non-absorbing in the wavelength range that we study. GaP has an indirect bandgap, with the stronger, direct optical transitions starting at λ ≈ 450 nm. Therefore, at λ = 500 nm, $L_{abs}$ shows a relatively large value of 10 μm, which decreases to 1 μm at λ = 450 nm and 100 nm at λ = 400 nm.

## Model for signal enhancement in signal-integration mode

We consider the following optical effects (see Figure 1 for a schematic and the Supporting Information for technical details): The enhancement of the excitation intensity at the location of the fluorophore, $\mathrm{ENH}_{exc}$, the enhancement in the fraction of emitted photons that are collected, $\mathrm{ENH}_{coll}$, the modification of the Purcell factor, $C_{\mathrm{Purcell}}$, and the parasitic absorption of emitted photons, $A_{\mathrm{parasitic}}$. Here, it is important to note that all these four optical properties depend on the materials and geometry of the nanowire system, as well as on the position of the fluorophore. $C_{\mathrm{Purcell}}$ modifies the quantum yield (QY) of the fluorophore, leading to a modification of the quantum yield, quantified by $\mathrm{ENH}_{QY} = C_{\mathrm{Purcell}}/(C_{\mathrm{Purcell}} \mathrm{QY}_0 + (1 - \mathrm{QY}_0))$ (see Supporting Information for details). Here, $\mathrm{ENH}_{exc}$, $\mathrm{ENH}_{coll}$, $C_{\mathrm{Purcell}}$, $\mathrm{QY}_0$, and $\mathrm{ENH}_{QY}$ are with reference to a fluorophore free in a homogeneous reference liquid with the same $n$ = 1.33 as for the liquid surrounding the nanowire.

$\mathrm{ENH}_{exc}$ depends on $\lambda_{exc}$ and the angle(s) from which the light is incident from. When light is incident from within $\mathrm{NA}_{exc}$, we calculate $\mathrm{ENH}_{exc}$ by incoherent integration over the angles within the $\mathrm{NA}_{exc}$ to represent the excitation in a widefield fluorescence microscope (see Supporting Information for details). Similarly, $\mathrm{ENH}_{coll}$ depends on $\lambda_{em}$ and the emission angles that are collected. When light is collected with a given $\mathrm{NA}_{coll}$, we calculate $\mathrm{ENH}_{coll}$ by integration of the emission over the angles within $\mathrm{NA}_{coll}$. Thus, in this signal-integration mode, we assume that all fluorescence light that enters $\mathrm{NA}_{coll}$ of the collection optics contributes to the detection signal.

$C_{\mathrm{Purcell}}$ and $A_{\mathrm{parasitic}}$ depend on $\lambda_{em}$, but not on $\lambda_{exc}$, and are independent of the excitation and collection optics used, and hence independent of $\mathrm{NA}_{exc}$ and $\mathrm{NA}_{coll}$.

We assume a case where we are far from excitation saturation, in which case $\mathrm{ENH}_{exc}$ modifies the rate of the excitation/de-excitation cycle and hence the number of emitter photons (see Supporting Information for details). In that case, the fluorescence signal enhancement is modelled as:

$$\mathrm{ENH}_{sig} = \mathrm{ENH}_{exc} \mathrm{ENH}_{coll} \mathrm{ENH}_{QY}, \qquad (1)$$

and $\mathrm{ENH}_{sig} = 1$ corresponds to equally strong signal as for the fluorophore in the homogenous liquid, for which the same $\mathrm{NA}_{exc}$ and $\mathrm{NA}_{coll}$ is used.

We denote by angled brackets averaging over the axial position of the fluorophore along the sidewall. The axially averaged signal-enhancement is thus given by $\langle \mathrm{ENH}_{sig} \rangle = \langle \mathrm{ENH}_{exc} \mathrm{ENH}_{coll} \mathrm{ENH}_{QY} \rangle$. In our study, $A_\mathrm{parasitic}$ affects $\mathrm{ENH}_\mathrm{sig}$ through $\mathrm{ENH}_\mathrm{coll}$ since parasitic absorption leads to a lower collection probability.

### Model for signal in imaging-mode

In the imaging mode, we consider the image that the fluorescence creates in an image plane, that is, when the far-field emission is focused with an imaging lens. Then, compared to the signal-integration mode above, we have here as an additional parameter $z_\mathrm{focal}$, the position of the focal plane of the imaging optics—see Figure 1(e) for a schematic. In our study, we calculate the 2D spatial distribution of the signal in the image plane (e.g. a camera chip, see Figure 1(e) for a schematic). For technical details of how the imaging modelling is performed, see the Supporting Information.

## Results

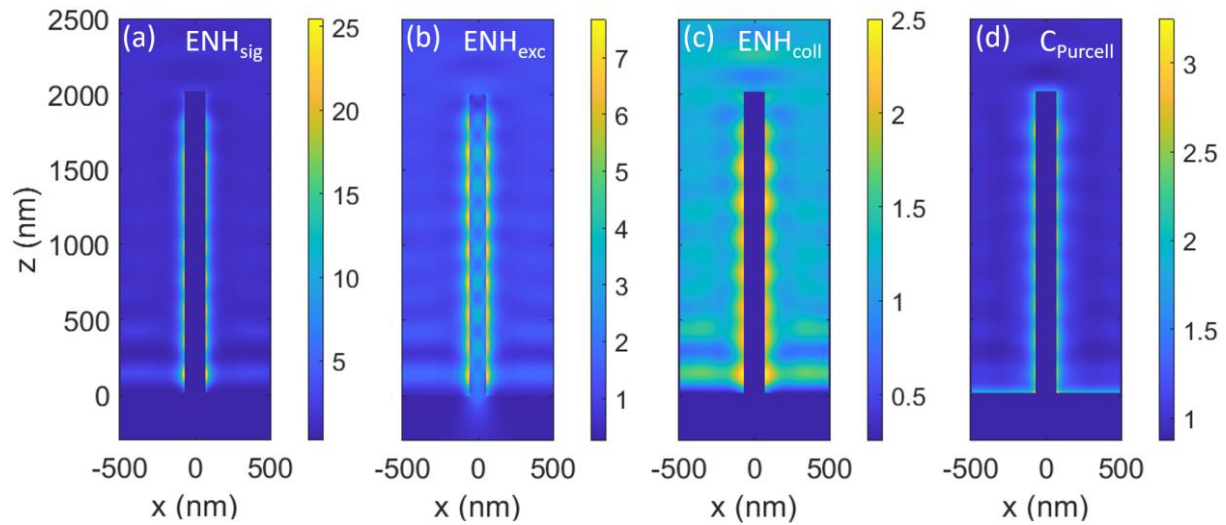

**Figure 2.** (a) Signal enhancement, (b) excitation enhancement, (c) collection enhancement, and (d) Purcell factor as a function of position around a GaP nanowire of *D* = 120 nm and *L* = 2000 nm, with a 10 nm thick SiO$_2$ coating and a GaP substrate. In (a) and (b), $\lambda_\mathrm{exc}$ = 640 nm and $\mathrm{NA}_\mathrm{exc} = 1$. In (a) and (c), $\lambda_\mathrm{em}$ = 670 nm and $\mathrm{NA}_\mathrm{coll} = 1$. In (d), $\lambda_\mathrm{em}$ = 670 nm. For (a), we assume $\mathrm{QY}_0 = 0.33$. These results are rotationally symmetric around the axis of the nanowire.

We show for *D* = 120 nm and *L* = 2000 nm in Figure 2 spatially resolved enhancement and Purcell factors for $\lambda_\mathrm{exc}$ = 640 nm, $\lambda_\mathrm{em}$ = 670 nm, and QY$_0$ = 0.33, which correspond to experiments with the AlexaFluor647 fluorophore.[27] This value for *D* optimizes the response for this combination of $\lambda_\mathrm{exc}$ and $\lambda_\mathrm{em}$, as detailed in the sections below. For all the factors in Figure 2, we find a strong localization to the surface of the nanowire, promoting signal from fluorophores bound to the nanowire as compared to a free fluorophores in the liquid, with signal enhancement within a region of approximately 100 nm from the nanowire sidewall (see Figure S4 in the Supporting Information). These values, obtained for $\mathrm{NA}_\mathrm{exc} = 1$ and $\mathrm{NA}_\mathrm{coll} = 1$, give after averaging over the nanowire length at the surface: $\langle \mathrm{ENH}_\mathrm{sig} \rangle =$ 16.0, $\langle \mathrm{ENH}_\mathrm{exc} \rangle = 4.8$, $\langle \mathrm{ENH}_\mathrm{coll} \rangle = 2.0$, and $\langle \mathrm{C}_\mathrm{Purcell} \rangle = 2.4$. Below, we investigate the different enhancement factors in detail, including a variation of QY$_0$, $\mathrm{NA}_\mathrm{exc}$, $\mathrm{NA}_\mathrm{coll}$, $\lambda_\mathrm{exc}$, and $\lambda_\mathrm{em}$, where especially by varying $\mathrm{NA}_\mathrm{exc}$ and $\mathrm{NA}_\mathrm{coll}$, more than two orders of magnitude enhancement in $\mathrm{ENH}_\mathrm{sig}$ can be reached. There, we focus on a fluorophore placed 5 nm on top of the side surface of the nanowire to represent a typical distance the fluorophore resides at.[26]

## Excitation enhancement

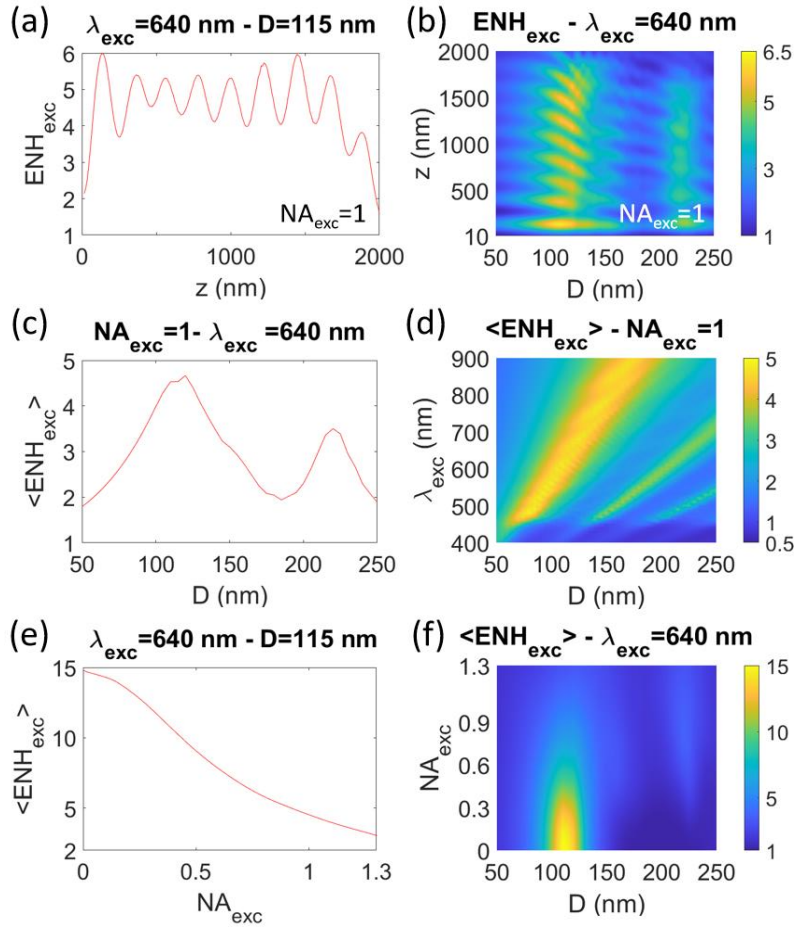

**Figure 3**. Enhancement of incident intensity for fluorophores at the side wall of GaP nanowires of $L$ = 2000 nm in length with a 10 nm thick $SiO_2$ coating. Dependence of $\text{ENH}_{\text{exc}}$ for $\lambda_{\text{exc}}$ = 640 nm as function of (a) axial position, with $z$ = 0 at the substrate surface and $z$ = 2000 nm at the top of the nanowire, for $\text{NA}_{\text{exc}} = 1$ and nanowire diameter $D$ = 115 nm, (c) as a function of $D$ for $\text{NA}_{\text{exc}} = 1$ after averaging along the axial position, that is, $\langle \text{ENH}_{\text{exc}} \rangle$, and (e) $\langle \text{ENH}_{\text{exc}} \rangle$ as a function of $\text{NA}_{\text{exc}}$ for $D$ = 115 nm. (b) Same as (a) but here additionally with diameter dependence. (d) Same as (c) but here additionally with wavelength dependence. (f) Same as (e) but here additionally with diameter dependence.

In Figure 3(a), we see $\text{ENH}_{\text{exc}} \approx 5$ for varying binding position $z$ along the nanowire axis for the excitation wavelength $\lambda_{\text{exc}}$ = 640 nm and $\text{NA}_{\text{exc}}$ = 1. There is a clear interference pattern due to the incident light and light reflected from the substrate interface—also seen in Figure 2(b). This standing wave pattern is evident also in Figure 3(b) that shows the dependence on $z$ and $D$. In Figure 3(f), it is seen that the strongest enhancement occurs around $\theta = 0$ for a nanowire diameter of $D$ = 115 nm where the $HE_{11}$ fundamental waveguide mode[25] is excited strongly. Thus, by decreasing $\text{NA}_{\text{exc}}$ we can increase $\langle \text{ENH}_{\text{exc}} \rangle$ to a value of approximately 15 from the value of approximately 5 at $\text{NA}_{\text{exc}} = 1$ (Figure 3(e,f)), in line with previous studies.[26] We note that the maximum is quite broad in wavelength in Figure 3(d) at a given $D$. Thus, even a rather broadband LED light source with 100 nm FWHM would cause only a minor decrease in the peak excitation enhancement. From Figure 3(c), it appears that we have a range of approximately ±10 nm around the optimum $D$ in which only a very minor decrease in $\langle \text{ENH}_{\text{exc}} \rangle$ occurs. The optimum diameter depends on $\lambda_{\text{exc}}$ and shifts almost linearly with $\lambda_{\text{exc}}$, as seen in Figure 3(d)—a linear shift would be expected[26] if the refractive indexes of the constituent materials were completely wavelength independent.

## Collection enhancement

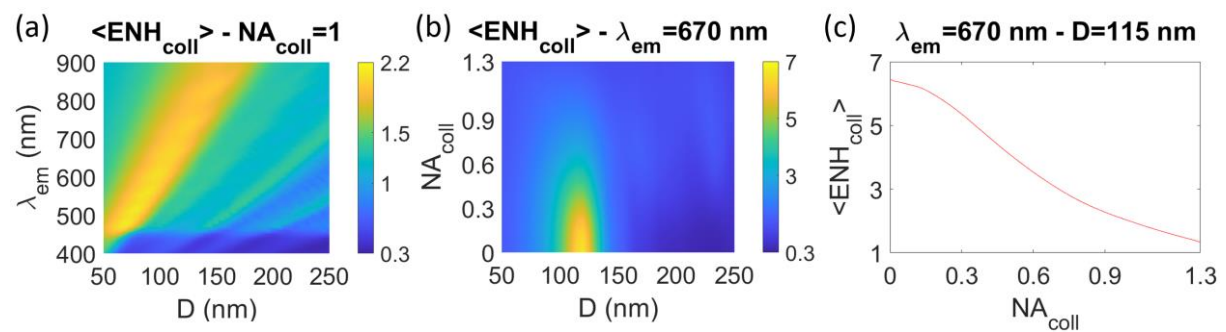

**Figure 4**. Enhancement of the collection of emission from fluorophores at the side wall of GaP nanowires of *L* = 2000 nm in length with a 10 nm thick SiO$_2$ coating as a function of (a) emission wavelength and nanowire diameter for a fixed $\text{NA}_{\text{coll}} = 1$ and (b) $\text{NA}_{\text{coll}}$ and nanowire diameter for a fixed emission wavelength of $\lambda_{\text{em}}$ = 670 nm. (c) Line-cut from (b) at *D* = 115 nm.

The collection enhancement in Figure 4(a) shows similar wavelength dependence as the excitation enhancement in Figure 3(d). However, for $\langle \text{ENH}_{\text{coll}} \rangle$, the optimum diameter at a given wavelength is shifted to a slightly smaller value and broadened. With $\text{NA}_{\text{coll}} = 1$, we reach $\langle \text{ENH}_{\text{coll}} \rangle \approx 2.0$ in Figure 4(a). The maximum in $\langle \text{ENH}_{\text{coll}} \rangle$ is quite broad in wavelength in Figure 4(a). Therefore, the effect from a moderately broadened emission spectrum of a fluorophore will be minor. Indeed, from the results in Figure 4(a), we can extract that the drop in the peak value of $\langle \text{ENH}_{\text{coll}} \rangle$ is on the order of 5% when the FWHM of the fluorescence spectrum increases to 100 nm.

By decreasing $\text{NA}_{\text{coll}}$, $\langle \text{ENH}_{\text{coll}} \rangle > 6.0$ can be reached (Figure 4(b,c)). However, such an increase in $\langle \text{ENH}_{\text{coll}} \rangle$ by decreasing $\text{NA}_{\text{coll}}$ will cause a drop in the actual signal, while increasing signal relative to the signal from the test fluorophore in the liquid, collected with the same $\text{NA}_{\text{coll}}$. In other words, such an increase in $\langle \text{ENH}_{\text{coll}} \rangle$ by decreasing $\text{NA}_{\text{coll}}$ will drop the actual signal while increasing signal-to-noise ratio if the dominating noise has optical origin and originates homogeneously from all angles within the $\text{NA}_{\text{coll}}$ of the collection objective. As an example, with the nanowire present, with $\text{NA}_{\text{coll}} = 0.1$, we collect 0.88% of the emitted photons, while with $\text{NA}_{\text{coll}} = 1.0$, we collect 34% (see Figure S5 in the Supporting Information). In the homogeneous liquid, with $\text{NA}_{\text{coll}} = 0.1$, we collect 0.14% of the emitted photons, while with $\text{NA}_{\text{coll}} = 1.0$, we collect 17% (see Figure S5 in the Supporting Information).

## Purcell factor and parasitic absorption

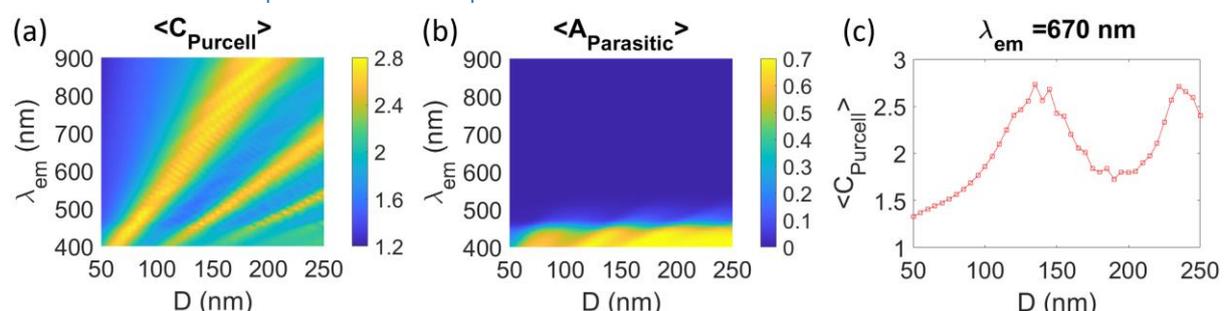

**Figure 5**. (a) Purcell factor and (b) parasitic absorption of emission from fluorophores at the side wall of GaP nanowires of *L* = 2000 nm in length with a 10 nm thick SiO$_2$ coating as a function of emission wavelength and nanowire diameter. (c) Line-cut from (a) at $\lambda_{\text{em}}$ = 670 nm.

We find that the Purcell factor is above unity for all diameter and emission wavelength values that we study (Figure 5(a))—see Figure 5(c) for a line-cut for $\lambda_{\text{em}}$ = 670 nm. $\langle C_{\text{Purcell}} \rangle$ peaks at $D \approx 140$ nm, which is shifted to a slightly larger value compared to the *D* = 115 nm that optimized $\langle \text{ENH}_{\text{coll}} \rangle$ for the same $\lambda_{\text{em}} = 670$ nm in Figure 4(b). By comparing Figure 5(a) and Figure 4(a), we find that the peaks in $C_{\text{Purcell}}$ are in general shifted to slightly larger diameter values compared to the peaks in $\langle \text{ENH}_{\text{coll}} \rangle$.

Noticeable parasitic absorption occurs for $\lambda < 450$ nm where the direct optical transitions in GaP start (see Figure S3 in the Supporting Information), and values of $\langle A_{\text{Parasitic}} \rangle > 0.6$ show up there (Figure 5(b)). For longer wavelengths, the parasitic absorption loss is minor due to the weak indirect optical transitions in GaP, and for $\lambda > 550$ nm, absorption seizes when entering the bandgap.

## Signal enhancement

If possible, the use of a high quantum yield fluorophore is advisable to increase the signal. However, in certain cases, only a low quantum yield fluorophore might fulfil requirements on excitation and emission wavelength range that needs to be used, or the stability, in terms of resilience against photobleaching, required for the detection. Thus, it is important to know the level of signal enhancement for different quantum yield of the fluorophore in the test liquid. Therefore, we consider here three values for $\text{QY}_0$. The values we consider are $\text{QY}_0 = 0_+, 0.33,$ and $1$. With $\text{QY}_0 = 0_+$ we denote a very low value of the quantum yield, approaching zero, in which case $\text{ENH}_{\text{QY}} = C_{\text{Purcell}}$. For a very high QY fluorophore, that is, when $\text{QY}_0 = 1$, $\text{ENH}_{\text{QY}} = 1$ and thus independent of $C_{\text{Purcell}}$.

We start by considering $\lambda_{\text{exc}} = 640$ nm and $\lambda_{\text{em}} = 670$ nm (Figure 6(a,c,d)). Such a detuning of 30 nm between excitation and emission wavelength, which can be accomplished by the choice of excitation wavelength for a given fluorophore emission wavelength, is motivated by the fact that each of excitation (that happens at $\lambda_{\text{exc}}$, Figure 3) and collection enhancement (that happens at $\lambda_{\text{em}}$, Figure 4) show a rather broad peak. Thus, with a 30 nm detuning, we expect to reach a rather optimized diameter that works well to enhance simultaneously excitation and collection. Indeed, we have found (Figure S6 in the Supporting Information) that we can allow approximately 50 nm detuning between $\lambda_{\text{exc}}$ and $\lambda_{\text{em}}$ before the peak value of $\langle \text{ENH}_{\text{sig}} \rangle$ starts to drop noticeably. Furthermore, a 30 nm detuning enables to easily use a long-pass filter in detection to filter out excitation light in experiments.[25–27]

With $\text{NA}_{\text{coll}} = \text{NA}_{\text{exc}} = 1$ (Figure 6(a)), we find for $\langle \text{ENH}_{\text{sig}} \rangle$ a peak value of 22.7 for $\text{QY}_0 = 0_+$, 15.5 for $\text{QY}_0 = 0.33$, and 9.4 for $\text{QY}_0 = 1$, all at *D* = 120 nm. With $\text{NA}_{\text{coll}} = 1$ and $\text{NA}_{\text{exc}} = 0.1$ (Figure 6(c)), we find for $\langle \text{ENH}_{\text{sig}} \rangle$ a peak value of 63.9 for $\text{QY}_0 = 0_+$ at *D* = 120 nm, 45.1 for $\text{QY}_0 = 0.33$ at *D* = 110 nm, and 29.1 for $\text{QY}_0 = 1$ at *D* = 110 nm. With $\text{NA}_{\text{coll}} = \text{NA}_{\text{exc}} = 0.1$ (Figure 6(d)), we find for $\langle \text{ENH}_{\text{sig}} \rangle$ a peak value of 213 for $\text{QY}_0 = 0_+$ at *D* = 120 nm, 147 for $\text{QY}_0 = 0.33$ at *D* = 115 nm, and 91.6 for $\text{QY}_0 = 1$ at *D* = 115 nm. Since the difference between $\langle \text{ENH}_{\text{sig}} \rangle$ for $\text{QY}_0 = 0_+$ and $\text{QY}_0 = 1$ is proportional to $C_{\text{Purcell}}$, we estimate from the difference in the respective peak values in Figure 6(a,c,d) a $C_{\text{Purcell}}$ of approximately 2.3, in good agreement with the values for $C_{\text{Purcell}}$ in Figure 5(c).

We have also extracted corresponding values for the fluorophore on the planar substrate surface, with 10 nm thick SiO$_2$ oxide as also used for the nanowire. For a SiO$_2$ substrate, we find $\langle \text{ENH}_{\text{sig}} \rangle$ in the range of 0.7 to 0.8 for all $\text{NA}_{\text{coll}}$, $\text{NA}_{\text{exc}}$, and $\text{QY}_0$ considered. Such a small difference from the unit value is expected since the refractive index of SiO$_2$ differs only by a minor amount of ≈0.1 from that of the *n* = 1.33 surrounding liquid. In contrast, for GaP substrate of considerably higher *n*, $\langle \text{ENH}_{\text{sig}} \rangle$ drops considerably, for example to 0.24 for $\text{NA}_{\text{coll}} = \text{NA}_{\text{exc}} = 1$ and $\text{QY}_0 = 0.33$ and to 0.11 for $\text{NA}_{\text{coll}} = \text{NA}_{\text{exc}} = 0.1$ and $\text{QY}_0 = 0.33$. Thus, for a fluorophore bound to the nanowire compared to the GaP substrate at $\text{NA}_{\text{coll}} = \text{NA}_{\text{exc}} = 0.1$ and $\text{QY}_0 = 0.33$, we expect a difference by a factor of 147/0.11 > $10^3$ in $\langle \text{ENH}_{\text{sig}} \rangle$, and for $\text{NA}_{\text{coll}} = \text{NA}_{\text{exc}} = 1$ and $\text{QY}_0 = 0.33$, we expect a difference by a factor of 15.5/0.24 > 60 in $\langle \text{ENH}_{\text{sig}} \rangle$.

As long as both the excitation and emission wavelength are above 450 nm where the strong absorption in GaP seizes, we can find a clear optimum diameter for varying emission wavelength, that is, for fluorophores emitting at different wavelength—see Figure 6(b) where a 30 nm detuning is assumed between $\lambda_{exc}$ and $\lambda_{em}$—the optimum diameter shows a rather linear shift with $\lambda_{em}$, as expected[26] since the constituent materials show only a minor wavelength dependence of their refractive indexes.

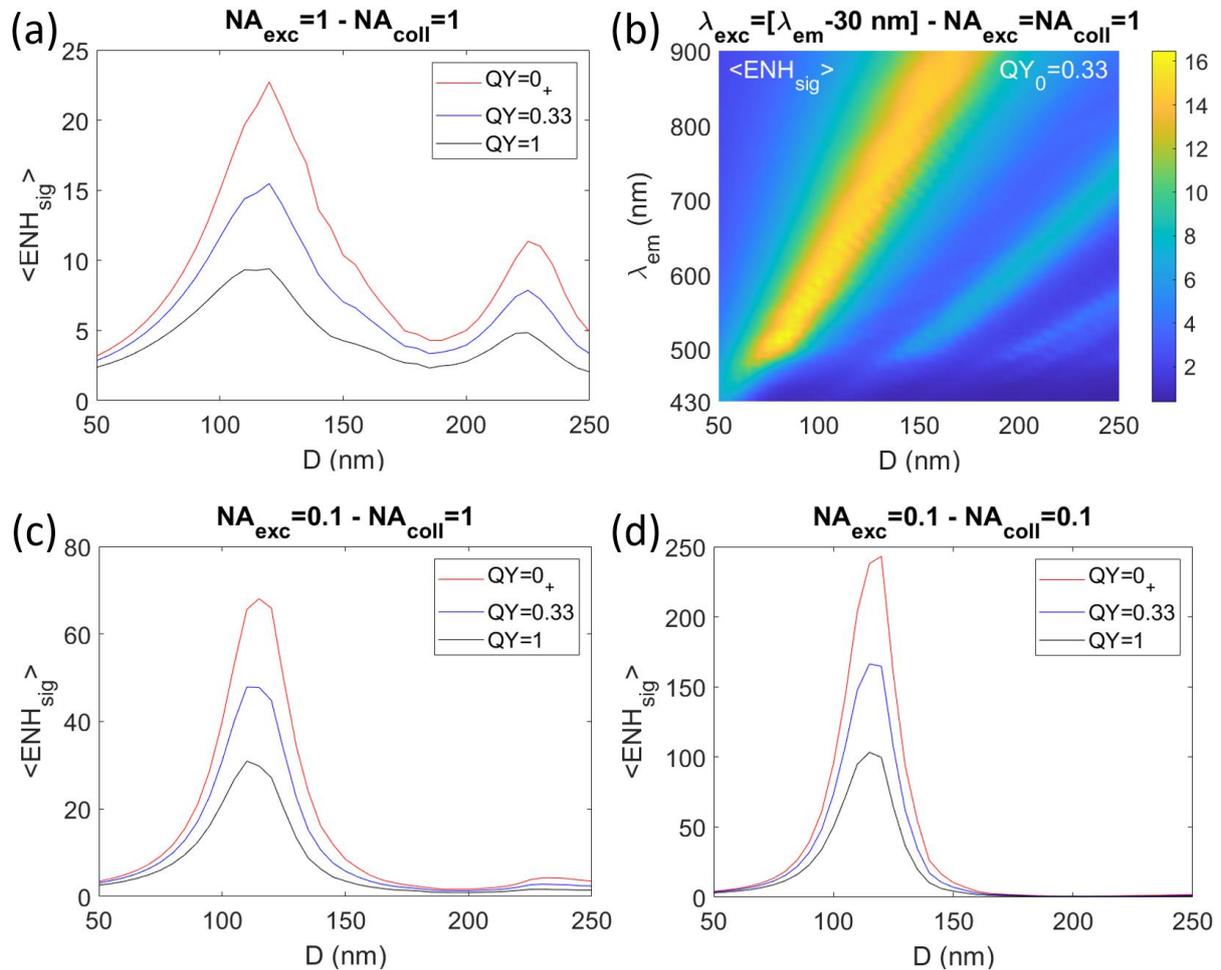

**Figure 6.** Signal enhancement as a function of nanowire diameter from fluorophores, averaged over axial position, that is, $\langle \mathrm{ENH}_{sig} \rangle$, for GaP nanowires of $L$ = 2000 nm in length with a 10 nm thick SiO$_2$ coating. Here, $\lambda_{exc}$ = 640 nm and $\lambda_{em}$ = 670 nm in (a), (c) and (d), and we show results for $\mathrm{QY}_0 = 0_+, 0.33$, and 1. (a) $\mathrm{NA}_{excl} = \mathrm{NA}_{coll} = 1$, (c) $\mathrm{NA}_{exc} = 0.1$ and $\mathrm{NA}_{coll} = 1$; and (d) $\mathrm{NA}_{exc} = \mathrm{NA}_{coll} = 0.1$. (b) $\mathrm{NA}_{exc} = \mathrm{NA}_{coll} = 1$, $\mathrm{QY}_0 = 0.33$, and the excitation wavelength is related to the emission wavelength through a 30 nm offset, that is, $\lambda_{exc} = \lambda_{em} - 30$ nm.

As seen in Figure 2, $\langle \mathrm{ENH}_{sig} \rangle$ drops quickly when moving away from the nanowire surface. A similar effect is obtained if coating the nanowire with a thicker SiO$_2$ layer. Compared to the $\langle \mathrm{ENH}_{sig} \rangle = 15.5$ obtained above for $\mathrm{QY}_0 = 0.33$ in Figure 6(a) for the oxide of thickness 10 nm, for oxide thickness of 0, 30, 50, and 70 nm, we find $\langle \mathrm{ENH}_{sig} \rangle$ = 22.0, 7.9, 4.7, 3.2, and 2.4, respectively (Figure S7 in Supporting Information). Thus, we recommend to aim for as thin oxide coating as possible, while still enabling biofunctionalization of the nanowire surface.

For the GaP nanowire, we find that for $\lambda_{exc}$ = 640 nm and $\lambda_{em}$ = 670 nm, that is, in a non-absorbing regime, the signal enhancement is rather independent of $L$, as long as $L$ > 1000 nm (Figure S8 in the

Supporting Information). Such a result is expected due to the rather dense fringe pattern along z (see Figure 2(a))—at large enough $L$, an increase in $L$ should then not cause a strong variation in $\langle \text{ENH}_{\text{sig}} \rangle$.

By varying the (real part of the) refractive index of the nanowire material from that of GaP, for $\lambda_{\text{exc}} = 640$ nm, $\lambda_{\text{em}} = 670$ nm, $\text{NA}_{\text{coll}} = \text{NA}_{\text{exc}} = 0.1$ and $\text{QY}_0 = 0.33$, with $\text{Re}(n_{\text{NW}})$ = 2.0, 2.5, 3.0, 3.5, 4.0, and 4.5, we find a maximum value of $\langle \text{ENH}_{\text{sig}} \rangle$ = 4.7, 8.2, 12.5, 17.6, 22.8 and 26.9 at $D$ = 180, 140, 130, 115, 100, and 90 nm, respectively (Figure S9 in the Supporting Information). Thus, the signal enhancement is stronger for higher refractive index of the nanowire, and the optimum diameter scales rather linearly with the inverse of $\text{Re}(n_{\text{NW}})$. By instead varying $\text{Im}(n_{\text{NW}})$ to induce absorption in the nanowire, we find that for a bulk absorption length of $L_{\text{abs}} = 2000$ nm, equal to the considered $L$ above, $\langle \text{ENH}_{\text{sig}} \rangle$ = 11.6, as compared to 15.5 for the case of no absorption (Supporting Information Figure S9 in the Supporting Information), decreasing to 6.0 at $L_{\text{abs}} = 500$ nm, and to 1.1 at $L_{\text{abs}} = 100$ nm.

These results on $\text{Re}(n_{\text{NW}})$ and $\text{Im}(n_{\text{NW}})$ explain the differences found between nanowires of different materials, shown in Figures S11-S17 in the Supporting Information, where we consider in addition to GaP also Si, GaAs, ZnO and GaN, with their refractive indexes shown in Figure S3 in the Supporting Information. These nanowire materials span a large range of refractive index values, with ZnO showing $\text{Re}(n) \approx 2.0$, GaN showing $\text{Re}(n) \approx 2.3$, and Si, GaAs, and GaP showing $\text{Re}(n)$ in the range from 3 to 5.5, with noticeably wavelength dispersion. Importantly, ZnO and GaN do not show absorption in this wavelength range. In contrast, Si, GaAs, and GaP show $\text{Im}(n) > 0$ with considerable wavelength dependence (Figure S3 in the Supporting Information). This $\text{Im}(n)$ translates into the bulk absorption length $L_{\text{abs}}$ shown in Figure S3 in the Supporting Information. Out of all the five nanowire materials studied, GaAs absorbs light strongest in this wavelength range thanks to its direct bandgap corresponding to $\lambda \approx 870$ nm—and for $\lambda > 870$ nm, the absorption is negligible. For GaAs, already at $\lambda \approx 850$ nm, $L_{\text{abs}}$ has dropped to 1 µm and continues to drop to 200 nm at $\lambda \approx 600$ nm and 14 nm at $\lambda \approx 400$ nm. The bandgap of Si is at $\lambda \approx 1100$ nm, beyond the range studied here. Therefore, in the full wavelength range studied, Si absorbs light. However, due to the indirect bandgap of Si, the absorption length stays rather long for $\lambda > 500$ nm: $L_{\text{abs}} = 10$ µm at $\lambda$ = 780 nm, $L_{\text{abs}} \approx 1$ µm at $\lambda$ = 500 nm, and $L_{\text{abs}} \approx 100$ nm at $\lambda$ = 400 nm. As an example, for $\lambda_{\text{exc}} = 640$ nm, $\lambda_{\text{em}} = 670$ nm, $\text{NA}_{\text{coll}} = \text{NA}_{\text{exc}} = 0.1$ and $\text{QY}_0 = 0.33$, Si, GaAs, ZnO, and GaN show 19.6, 8.5, 3.6, and 5.9, respectively, (Figure S11-S17 in the Supporting Information), as compared to the 15.5 for GaP.

### Detection in imaging mode

We are consider the image sharpness around the position of the nanowire in the image (Figure 1(e)).[27] Therefore, we focus on how much of the overall intensity captured by the collection objective falls within a disc of radius $r_{\text{img}}$ centered on the nanowire. In other words, we are interested in the integrated intensity on the pixels within a disc of radius $r_{\text{img}}$ relative to the integrated intensity on all pixels—we denote this fraction $\text{REL}_{\text{img}}$. We also consider how much more counts we observe in that imaging disc, relative to the counts in that imaging disc for a fluorophore, in best focus, in the homogeneous test liquid of $n$ = 1.33—we call this ratio $\text{ENH}_{\text{img}}$, and with $r_{\text{img}} \to \infty$, $\text{ENH}_{\text{img}} \to \text{ENH}_{\text{sig}}$.

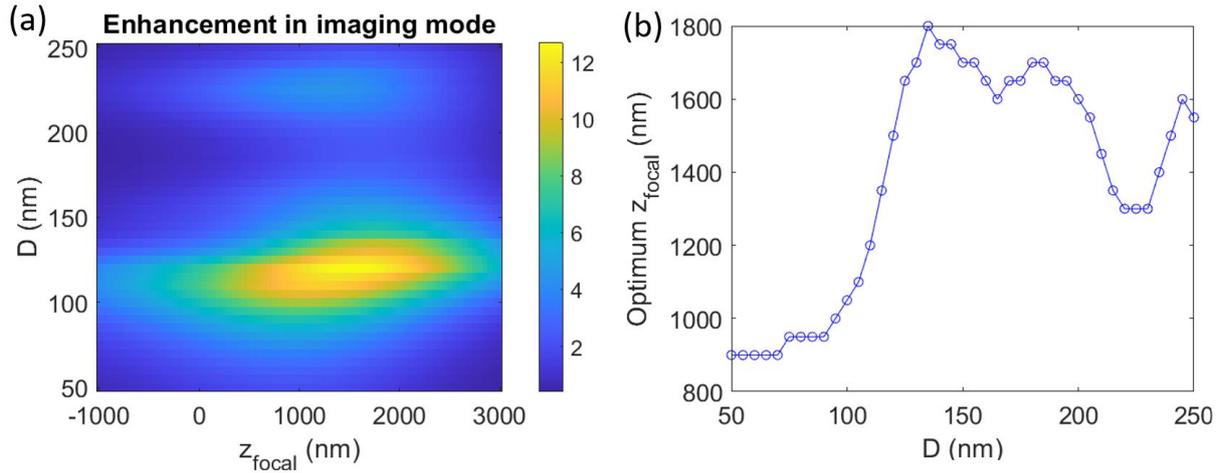

**Figure 7.** (a) Enhancement in imaging mode, $\langle \text{ENH}_{\text{img}} \rangle$, for $r_{\text{img}} = 500$ nm. (b) Optimum $z_{\text{focal}}$ from (a), and the discrete-appearing nature of the graph is due to the 50 nm step used in the modelling of $z_{\text{focal}}$. Here, we consider GaP nanowires of $L$ = 2000 nm in length with a 10 nm thick SiO$_2$ coating, $\lambda_{\text{exc}} = 640$ nm, $\lambda_{\text{em}} = 670$ nm, $QY_0 = 0.33$, and $\text{NA}_{\text{exc}} = \text{NA}_{\text{coll}} = 1$.

We find that in imaging mode, with the use of the $r_{\text{img}} = 500$ nm, we reach almost as good enhancement (82% of it) as in the non-imaging detection mode (Figure 7(a) and Figure S18 in the Supporting Information), while if we decrease $r_{\text{img}}$ to 250 nm, we reach 65% of it—a detailed investigation of the effect of $r_{\text{img}}$ on $\text{ENH}_{\text{img}}$ is left for a future study. Further, we find that the optimum focal plane at the small diameter values of D < 100 nm is at approximately $L/2$ (Figure 7(b)), as expected for nanowires that are not lightguiding,[25] in which case we expect to obtain the best result by placing the focal plane at the centre of the out-of-plane distribution of fluorophores (Supporting Information Figure S19 the Supporting Information). The rapid increase in optimum $z_{\text{focal}}$ at $L$ > 100 nm in Figure 7(b) is an indication of light-guiding. Indeed, for $D$ = 120 nm, which optimizes the signal in Figure 7(a), with varying nanowire length, we find that the optimum focusing plane is approximately 300-400 nm below the top of the nanowire (Figure S20 the Supporting Information). For this value of $D$, we do find both a light-guiding and non-lightguiding component in the emission (Figure S21 in the Supporting Information). Thanks to the strong light-guiding component, we can focus to close to the top of the nanowire and still receive a sharp image, irrespective of the actual binding position of the fluorophore along the nanowire (Figure S21 the Supporting Information), as found also in recent experiments.[24,25,28]

The light-guiding is seen clearly in $\text{REL}_{\text{img}}$ when considering $L$ = 4000 nm at the optimum $z_{\text{focal}}$ = 3750 nm (Figure S22 the Supporting Information). There, even from the bottom of the nanowire, ≈50% of the photons in the image plane are in the 500 nm radius detection disc. In contrast, for the case of the homogeneous liquid, at a 3700 nm offset between the fluorophore and the focal plane, approximately 2% of the emitted photons are within that detection disc (Figure S19 in the Supporting Information).

As another effect from light-guiding, for a focus offset of 2000 nm, $\text{REL}_{\text{img}} \approx 5\%$ for a fluorophore on planar GaP substrate (Figure S19 in the Supporting Information), relative to which the signal itself is enhanced by a factor of 15.5/0.24 ≈ 60 for the nanowire-bound fluorophore (as discussed above). Thus, the image of the flurophore is blurred out by the relative factor of more than 0.5/0.05 = 10, and on top of that, decreased overall in intensity by the factor 60, explaining why fluorophores from the substrate surface appear to give negligible contribution in recent imaging experiments when focusing toward the top of the nanowire.[27,29]

Further analysis has shown that if we shift to consider $\text{NA}_{\text{coll}} = 0.1$ instead of $\text{NA}_{\text{coll}} = 1$, the imaging mode results follow the results for the non-imaging detection mode very closely. The focal depth in homogeneous medium is in this case, for $\lambda_{\text{em}}$ = 670 nm, considerably larger than the *L* = 2000 nm. In that case, the exact focal plane position along the nanowire length has a negligible effect on image formation. As an example, for the homogeneous reference system, it takes a 40 µm shift of the focal plane to cause a drop by 10% (relative) of the counts in a detection disc of 500 nm in radius in the image plane, as compared to in best focus. At the same time, the image is blurred out strongly when reducing $\text{NA}_{\text{coll}}$. For example, for the homogeneous reference system at best focus, for $\text{NA}_{\text{coll}} = 0.1$, just 5% of the counts are within the disc of $r_{\text{img}} = 500$ nm. Thus, if using $\text{NA}_{\text{coll}} = 0.1$ to enhance signal-to-noise ratio, we are inclined to opt for a simple non-imaging detection scheme since additional benefits from imaging detection could be questionable.

## Conclusions

For material choice, we found that a higher refractive index for the nanowire material is beneficial for the signal enhancement. From that perspective, GaP, Si, and GaAs are a better choice than ZnO and GaN (Figure S3 in the Supporting Information). However, parasitic absorption in the nanowire material can also be a limiting factor. Due to this, Si is recommended for wavelengths beyond approximately 650 nm and GaAs for wavelengths beyond 870 nm, while GaP appears suitable for wavelengths beyond 450 nm (Figure S3 in the Supporting Information)—results that we acquired for 2000 nm long nanowires.

Regarding geometry, to increase the binding area for the fluorophore, we could use a larger value for *D*. In that case, a fluorophore with longer emission wavelength appears more optimum (Figure 6b). On the other hand, the binding area can be increased also by increasing *L* at fixed *D*, in which case the choice for the fluorophore would not need to be changed (Figure S8 in the Supporting Information). Throughout, we have analysed a single nanowire, in line with our recent experiments on sparse nanowire arrays where nanowire-to-nanowire optical coupling is negligible.[27] For future studies, analysis of the optical effects from closer packed nanowires would be valuable to aid in the array design.

We showed how by decreasing both $\text{NA}_{\text{exc}}$ and $\text{NA}_{\text{coll}}$ (Figure 6(a,c,d)), it is possible to enhance the signal from a fluorophore bound to the nanowire by a factor of >$10^2$ relative to a fluorophore in homogenous test liquid and by >$10^3$ relative to a fluorophore on high refractive index GaP substrate. At the same time, it is important to keep in mind that, at optimized diameter, a decrease in $\text{NA}_{\text{exc}}$ does not cause a drop in signal while a decrease in $\text{NA}_{\text{coll}}$ causes a drop in the actual signal (Figure S5 in the Supporting Information). Thus, limiting of both $\text{NA}_{\text{exc}}$ and $\text{NA}_{\text{coll}}$ is a valuable strategy if signal-to-noise ratio is limited by autofluorescense that is not directional. On the other hand, limiting of just $\text{NA}_{\text{exc}}$ is a valuable approach if it is the actual signal that limits the detectivity and the optical excitation power cannot or should not be increased, for example due to heating of the sample and/or autofluorescence that increases superlinearly with excitation power.

Finally, the signal enhancement by the factor of 213 for a low quantum yield fluorophore and 92 for a high quantum yield fluorophore in Figure 6(d) for $\mathrm{NA_{coll}} = \mathrm{NA_{exc}} = 0.1$ is very promising for signal detection with a small numerical aperture (NA) for the combined excitation and detection system. Thus, the nanowire-based biosensor offers the prospect of high-signal biosensing with a portable, low-cost low-NA optical read-out system, even out in the field, i.e., in a lab-on-a-chip system.[1] Here, it should be remembered that the signal enhancement factor is relative to performing the detection with the same $\mathrm{NA_{coll}} = \mathrm{NA_{exc}} = 0.1$ for the fluorophore in the homogeneous liquid. It might be more fair to compare to the case of using a higher $\mathrm{NA_{coll}}$ of for example 1.0 in the homogeneous test liquid. In that case, the signal collected in the test liquid increases by a factor of 121 (Figure S5 in the Supporting Information). Thus, the actual signal with the low NA-optics with the nanowire biosensor increases by a factor or 213/121 ≈ 1.8 and 92/121 ≈ 0.75 for the low and high QY fluorophore, compared to using the high NA-optics for the case of the fluorophore in the test liquid. Thus, the low-NA optics with nanowire would yield comparable, or even higher, actual signal than with a high-NA non-nanowire conventional platform.

## Acknowledgement

We acknowledge financial support from the Waldemar von Frenckell foundation, Åbo Akademi University Foundation (Gösta Branders research fund), and the Research Council of Finland project 359066 (HPC-Phot). The computer resources of the Finnish IT Center for Science (CSC) and the FGCI project (Finland) are acknowledged. We acknowledge discussions about the nanowire biodetector with the Heiner Linke group, the Fredrik Höök group, the Christelle Prinz Group, and the Thoas Fioretos group. We are especially thankful for the discussions with Rubina Davtyan about the optics modelling and optical response of the nanowire biosensor.

## Disclosures

N.A. has financial interests in AlignedBio AB, Sweden.

## Data availability

The data from this study is provided in the manuscript, Supplementary Information figures and from the corresponding authors upon request.

# Supporting information

# Fluorophore signal detection and imaging enhancement in high refractive index nanowire biosensors

Nicklas Anttu[1]

[1]Physics, Faculty of Science and Engineering, Åbo Akademi University, FI-20500 Turku, Finland

Author to whom correspondence should be addressed: nicklas.anttu@abo.fi

# Contents

## Electromagnetic description of excitation and emission

We assume time-harmonic fields, and we solve for the **E** and **H** fields in complex notation with the convention for time-dependence of the form $\exp(-i\omega t)$ where $\omega = 2\pi c/\lambda$ with $c$ the speed of light in vacuum and $\lambda$ the (vacuum) wavelength.[1]

We assume dipole-type excitation and emission of the fluorophore. Furthermore, we assume an inherently isotropic dipole. Then, the excitation probability of the fluorophore is given by the $|\mathbf{E}|^2$ of the incident light at the location of the fluorophore. Similarly, the emission and hence emission modification of the dipole can be modelled by an electrical dipole, represented by a dipole moment **p**, and we average over *x*-, *y*-, and *z*-oriented dipole moment for a given fluorophore position to mimic an isotropic dipole. Note that the generalization to a non-isotropic underlying optical response of the fluorophore is straight-forward to take into account for the excitation light by taking appropriately weighted average of $|E_x|^2$, $|E_y|^2$, and $|E_z|^2$ of the excitation light at the location of the dipole, with the weighting given by the *x*-, *y*-, and *z*-component of the dipole moment of the fluorophore at the excitation wavelength. Similarly, we can take into account in emission a non-isotropic dipole by appropriate weighting of the response of *x*-, *y*-, and *z*-oriented dipole moment at the emission wavelength.[2]

## Modelling of Purcell factor and parasitic absorption

We perform individual modelling for *x*-, *y*-, and *z*-oriented dipole emitter at a given position, with dipole moment strength *p*. For each orientation of the dipole moment, in our simulations, we integrate the Poynting vector through the surface of a sphere of 4 nm in radius, centered at the dipole position. This integration yields $P_{em,i}$, the power emitted by the dipole. Here, *i* indicates the orientation of the dipole moment. Then, the Purcell factor is defined as:

$$C_{\text{Purcell}} = \frac{P_{em,x}+P_{em,y}+P_{em,z}}{P_{ref}} = \frac{P_{tot}}{P_{ref}}. \tag{S1}$$

Here, $P_{ref} = 3P_{homog.dip}$ with $P_{homog.dip}$ the emission power from a dipole in a homogenous surrounding of the same refractive index as of the material in which the dipole is placed (i.e., the liquid surrounding the nanowires in our case), and $P_{homog.dip}$ can be calculated analytically.[3]

For the modelling of the parasitic absorption in the nanowire, we extract for each modelled dipole orientation $P_{abs,i}$, the power absorbed in the nanowire (which we obtain as a volume integration of the spatially resolved absorption in the nanowire volume). Then, $A_{\text{Parasitic}}$ is the fraction of emitted light that is parasitically absorbed, and it is obtained by calculating the probability that emitted light from a certain dipole orientation is parasitically absorbed and multiplying with the probability that the emission occurs through that dipole orientation:

$$A_{\text{Parasitic}} = \frac{P_{abs,x}}{P_{em,x}} \times \frac{P_{em,x}}{P_{tot}} + \frac{P_{abs,y}}{P_{em,y}} \times \frac{P_{em,y}}{P_{tot}} + \frac{P_{abs,z}}{P_{em,z}} \times \frac{P_{em,z}}{P_{tot}} = \frac{P_{abs,x}+P_{abs,y}+P_{abs,z}}{P_{tot}}. \tag{S2}$$

Due to the circular symmetry of the system, we have only *r* and *z* dependence in $C_{\text{Purcell}}$ and $A_{\text{Parasitic}}$ (in addition to $\lambda$ dependence).

In our simulations of the nanowire, the dipole emitter is placed 5 nm from the oxide surface (except those shown in Figure 2 where also larger distance is considered), and therefore the sphere of 4 nm in radius in the calculation of $P_{em,i}$ is in a non-absorbing region. We perform the simulations with a step of 50 nm in the *z* position of the dipole, with linear interpolation of the results for positions between the modelled positions.

## Modelling of excitation enhancement

For the excitation, we assume an incoherent angular spectrum to resemble the illumination in a wide-field microscope in the Köhler illumination configuration. In other words, we assume that the incident light can be described as an incoherent superposition of plane waves. Thus, similarly as in Ref. [4], for the wide-field illumination, we assume incoherent plane waves at the excitation wavelength $\lambda_{\text{exc}}$ from within the excitation numerical aperture $\text{NA}_{\text{exc}}$, with $\theta_{\text{NA,exc}} = \arcsin\left(\frac{\text{NA}_{\text{exc}}}{n_{\text{inc}}}\right)$ the maximum incident angle and $n_{\text{inc}}$ the refractive index of the material on the incidence side (i.e., the liquid surrounding the nanowire in our simulations). Then, the enhancement of the excitation intensity, relative to the case of a fully homogenous medium, at location **r** is given by:

$$\text{ENH}_{\text{exc}}(\mathbf{r}, \lambda_{\text{exc}}) = \frac{\sum_{pol=s,p} \int_0^{\theta_{\text{NA,exc}}} \int_0^{2\pi} |\mathbf{E}(\mathbf{r}, \lambda_{\text{exc}}, \theta_{\text{inc}}, \phi_{\text{inc}}, pol)|^2 \sin(\theta_{\text{inc}}) d\phi_{\text{inc}} d\theta_{\text{inc}}}{\sum_{pol=s,p} \int_0^{\theta_{\text{NA,exc}}} \int_0^{2\pi} (\text{E}_{\text{inc}})^2 \sin(\theta_{\text{inc}}) d\phi_{\text{inc}} d\theta_{\text{inc}}}. \quad \text{(S3)}$$

Here, $\text{E}_{\text{inc}}$ is the electric field strength of the incident plane waves, which we set independent of incidence angle and polarization. $\mathbf{E}(\mathbf{r}, \lambda_{\text{exc}}, \theta_{\text{inc}}, \phi_{\text{inc}}, pol)$ is the electric field strength at position **r** caused by an incident plane wave from a direction given by polar angle $\theta_{\text{inc}}$ and azimuth angle $\phi_{\text{inc}}$ for polarization *pol* (in our simulations, we use *s* and *p* polarization, but any set of two orthogonal polarization states should work). Importantly, when modelling $\mathbf{E}(\mathbf{r}, \lambda_{\text{exc}}, \theta_{\text{inc}}, \phi_{\text{inc}}, pol)$, we include a fully coherent description of the light from that incidence angle, but the contribution from two different angles, or polarization states, is taken into account in an incoherent manner where it is thus the $|\mathbf{E}|^2$-contributions that are added/integrated in Eq. (S3).

Therefore, $\text{ENH}_{\text{exc}}$ is the enhancement relative to the excitation intensity in the system with homogeneous liquid, that is, without nanowire, oxide, or substrate present. For our case of the circular symmetric system, with the illumination from all azimuth angles, we have dependence on just *r* and *z* in $\text{EXC}_{\text{enh}}$.

In the simulations for the excitation enhancement, we consider a region of 10 nm in extent in the radial direction on top of the oxide layer, and we average the excitation enhancement from this region to yield the *z*-dependent excitation enhancement value. We extract values with a 10 nm step in *z*, and we use a 5° step in the simulation of $\theta_{\text{inc}}$. Since we integrate in Eq. (S3) over all azimuth angles, a single simulation in azimuth angle is sufficient if we from that result instead integrate over the azimuth angle for position (for details, see the Supporting Information in Ref. [4]).

## Modelling of collection enhancement

For the calculation of the collection enhancement for emission at wavelength $\lambda_{\text{em}}$, we use Lorentz reciprocity[1,3,5] to yield information of the enhancement of the emission to within the collection numerical aperture $\text{NA}_{\text{coll}}$, with $\theta_{\text{NA,coll}} = \arcsin\left(\frac{\text{NA}_{\text{coll}}}{n_{\text{coll}}}\right)$ the maximum collection angle and $n_{\text{coll}}$ the refractive index of the material on the collection side (in our modelling, we consider collection on the same side as the excitation, and then $n_{\text{coll}} = n_{\text{inc}}$). With the Lorentz reciprocity, the information of how an incident plane wave couples to the position of a dipole reveals how the dipole emits in the reciprocal direction from which the plane wave is incident.[1,3,5] Furthermore, when we consider a dipole emitter with isotropic underlying optical response, the averaged emission from all three dipole moment orientations to a given polarization state in emission is proportional to $|\mathbf{E}(\mathbf{r}, \lambda_{\text{em}}, \theta_{\text{inc}}, \phi_{\text{inc}}, pol)|^2$.[1,3,5]

Then, the collection enhancement, relative to the case of a fully homogenous medium, at location **r** is given by:

$$\text{ENH}_{\text{coll}}(\mathbf{r}, \lambda_{\text{em}}) = \left[ \sum_{pol=s,p} \int_0^{\theta_{\text{NA,coll}}} \int_0^{2\pi} |\mathbf{E}(\mathbf{r}, \lambda_{\text{em}}, \theta_{\text{inc}}, \phi_{\text{inc}}, pol)|^2 \sin(\theta_{\text{inc}}) d\phi_{\text{inc}} d\theta_{\text{inc}} \right. /$$
$$\left. \sum_{pol=s,p} \int_0^{\theta_{\text{NA,coll}}} \int_0^{2\pi} (\text{E}_{\text{inc}})^2 \sin(\theta_{\text{inc}}) d\phi_{\text{inc}} d\theta_{\text{inc}} \right] \times \frac{1}{C_{\text{Purcell}}(\mathbf{r}, \lambda_{\text{em}})}. \qquad (S4)$$

Here, the normalization with the Purcell factor corrects for the fact that the total emitted power scales with the Purcell factor, and the Lorentz reciprocity theorem includes[1,3,5] this enhancement of total emitted power. With this definition for $\text{ENH}_{\text{coll}}$, the effect of parasitic absorption in the nanowire is included implicitly through $\text{ENH}_{\text{coll}}$, and $A_{\text{Parasitic}}$ can be used for separately analyzing how much of the emitted light is lost due to parasitic absorption in the nanowire.

In more detail, $\left[ \frac{\sum_{pol=s,p} \int_0^{\theta_{\text{NA,coll}}} \int_0^{2\pi} |E_i(\mathbf{r}, \lambda_{\text{em}}, \theta_{\text{inc}}, \phi_{\text{inc}}, pol)|^2 \sin(\theta_{\text{inc}}) d\phi_{\text{inc}} d\theta_{\text{inc}}}{\sum_{pol=s,p} \int_0^{\theta_{\text{NA,coll}}} \int_0^{2\pi} (\text{E}_{\text{inc}})^2 \sin(\theta_{\text{inc}}) d\phi_{\text{inc}} d\theta_{\text{inc}}} \right]$ with $i = x, y, z$ yields the power emitted from a dipole with $i$ oriented dipole moment in the vicinity of the nanowire into the collection objective, normalized to the summed power emitted from three orthogonal dipoles in the homogeneous surrounding. We call this fraction $\frac{P_{em,i,\text{NA-coll}}}{\sum_j P_{ref,j,\text{NA-coll}}}$. Thus, Eq. (S4) is equivalent to

$$\frac{\sum_i P_{em,i,\text{NA-coll}}}{\sum_j P_{ref,j,\text{NA-coll}}} \times \frac{1}{C_{\text{Purcell}}} = \frac{\sum_i P_{em,i,\text{NA-coll}}}{\sum_j P_{ref,j,\text{NA-coll}}} \times \frac{3 P_{homog.dip}}{[P_{em,x} + P_{em,y} + P_{em,z}]} = \frac{\sum_i P_{em,i,\text{NA-coll}}}{[P_{em,x} + P_{em,y} + P_{em,z}]} / \left[ \frac{\sum_j P_{ref,j,\text{NA-coll}}}{3 P_{homog.dip}} \right]$$

$$= \left[ \sum_i \frac{P_{em,i,\text{NA-coll}}}{P_{em,i}} \frac{P_{em,i}}{[P_{em,x} + P_{em,y} + P_{em,z}]} \right] / \left[ \frac{\sum_j P_{ref,j,\text{NA-coll}}}{P_{homog.dip}} \frac{P_{homog.dip}}{3 P_{homog.dip}} \right],$$ which is indeed the fraction of emitted power propagating into the collection objective from three orthogonally oriented dipole moments in the vicinity of the nanowire, normalized to the fraction of emitted power propagating into the collection objective from three orthogonally oriented dipole moments in the homogenous surrounding (the first fraction in both the denominator and nominator denotes the fraction of emitted power of a specific dipole moment orientation that propagates into the collection objective; and the second fraction denotes the probability that the emission occurs through that dipole moment orientation).

For our case of the circular symmetric system, with collection to all azimuth angles, we have dependence on just *r* and *z* in $\text{ENH}_{\text{coll}}$.

In the simulations for the collection enhancement, we consider a region of 10 nm in extent in the radial direction on top of the oxide layer, and we average the collection enhancement from this region to yield the *z*-dependent excitation enhancement value. We extract values with a 10 nm step in *z*, and we use a 5° step in the simulation of $\theta_{\text{inc}}$. Since we integrate in Eq. (S4) over all azimuth angles, a single simulation in azimuth angle is sufficient if we from that result instead integrate over the azimuth angle for position (for details, see the Supporting Information in Ref. [4])

## Modification of the quantum yield and de-excitation rate

We assume a simple model for the fluorophore de-excitation with two competing pathways, one radiative, which causes the emission of the photons that contribute to the detected signal, with rate $\Gamma_{\text{rad}}$ and one non-radiative with rate $\Gamma_{\text{nr}}$. Thus, the total de-excitation rate is $\Gamma = \Gamma_{\text{rad}} + \Gamma_{\text{nr}}$, the de-excitation time constant is given by $\tau = 1/\Gamma$, and the probability for emission of a photon during de-excitation is given by $\text{QY} = \Gamma_{\text{rad}}/\Gamma = \Gamma_{\text{rad}}/(\Gamma_{\text{rad}} + \Gamma_{\text{nr}})$. This simple two-pathway model should be applicable for realistic fluorophores if the following assumptions are fulfilled: (i) The internal conversion within the fluorophore occurs considerably faster than the final radiative or non-radiative decay and (ii) re-excitation, after non-radiate decay, from an intermediate energy state back to the excited state, which could give rise to delayed fluorescence, is negligible.[6] (If we model multiple fluorophores present with the use of this model, we implicitly assume that the excitation of a fluorophore by a photon emitted by another fluorophore can be neglected.)

As described above, for $C_{\text{Purcell}}$, we use a dipole model where we average over *x*-, *y*-, and *z*-oriented dipole moment to represent emission from an inherently unpolarized fluorophore and normalize to the emission in the test liquid, which in our case has $n = 1.33$. Thus, we obtain a relative Purcell factor with reference to the test liquid medium, with this reference value a factor of $n$ higher than that of vacuum.[2] We choose to work with this relative Purcell factor throughout since the QY and decay lifetime of fluorophores are typically tabulated in such a test liquid, and hence, as discussed below, modified by the relative Purcell factor.

We assume that the fluorophore in the reference liquid, where thus $C_{\text{Purcell}} = 1$, shows a QY of $\text{QY}_0$ and $\tau$ of $\tau_0$ (thus, $\Gamma_0 = 1/\tau_0$). The Purcell factor modifies $\Gamma_{\text{rad}}$ such that $\Gamma_{\text{rad}} = C_{\text{Purcell}}\Gamma_{\text{rad},0}$,[1] and for simplicity, we assume that the non-radiative pathway stays unaffected when the fluorophore is brought to the vicinity to the nanowire (and possibly bound to it by molecular linking). Then, it follows that $\text{QY} = C_{\text{Purcell}}\text{QY}_0/(C_{\text{Purcell}}\text{QY}_0 + (1 - \text{QY}_0))$. Thus, we can define a QY enhancement factor as $\text{QY} = \text{ENH}_{\text{QY}}\text{QY}_0$ where $\text{ENH}_{\text{QY}} = C_{\text{Purcell}}/(C_{\text{Purcell}}\text{QY}_0 + (1 - \text{QY}_0))$. Similarly, the de-excitation time constant is modified to $\tau = \frac{\tau_0}{C_{\text{Purcell}}\text{QY}_0 + (1-\text{QY}_0)}$, and we can define a de-excitation speed-up factor as $\text{ENH}_{1/\tau} = C_{\text{Purcell}}\text{QY}_0 + (1 - \text{QY}_0)$. Note that since $\text{ENH}_{\text{QY}}$ and $\text{ENH}_{1/\tau}$ depend here only on the Purcell factor, they are independent of $\text{NA}_{\text{exc}}$ and $\text{NA}_{\text{coll}}$.

## Model for fluorescence signal enhancement in signal-integrating collection

Here, we assume a signal-integrating collection scheme where all photons propagating within $\text{NA}_{\text{coll}}$ contribute to the fluorescence signal.

We consider two limiting cases for the excitation/de-excitation cycle: (1) Far-from saturation regime where the fluorophore is most of the time in the ground state, so that the excitation process limits the rate of the excitation/de-excitation cycle and hence signal. (2) The saturation regime where the fluorophore is most of the time in the excited state, and the signal is hence limited by the de-excitation rate $1/\tau$.

Then, for case (1), that is, far from excitation saturation, the signal enhancement is given by

$$\text{ENH}_{\text{sig}} = \text{ENH}_{\text{exc}}\text{ENH}_{\text{coll}}\text{ENH}_{\text{QY}}. \tag{S5}$$

Here, two special cases follow, one for fluorophores with low $\text{QY}_0$ (close to zero) where $\text{ENH}_{\text{QY}} \to C_{\text{Purcell}}$ and one for high $\text{QY}_0$ (close to 1) where $\text{ENH}_{\text{QY}} \to 1$. In this way, we obtain $\text{ENH}_{\text{sig}} = \text{ENH}_{\text{exc}}\text{ENH}_{\text{coll}}C_{\text{Purcell}}$ and $\text{ENH}_{\text{sig}} = \text{ENH}_{\text{exc}}\text{ENH}_{\text{coll}}$ for these limiting cases of $\text{QY}_0$, respectively.

For case (2), that is, in the saturation regime, the effect of $\text{ENH}_{\text{exc}}$ disappears, and instead $\text{ENH}_{1/\tau}$ modifies the excitation/de-excitation rate: $\text{ENH}_{\text{sig}} = \text{ENH}_{1/\tau}\text{ENH}_{\text{coll}}\text{ENH}_{\text{QY}}$. However, $\text{ENH}_{1/\tau}\text{ENH}_{\text{QY}} = C_{\text{Purcell}}$ and thus $\text{ENH}_{\text{sig}} = \text{ENH}_{\text{coll}}C_{\text{Purcell}}$ in this regime.

In the main text, we assume case (1), that is, the case far from excitation saturation.

Note that if using a broadband excitation source, we should average $\text{ENH}_{\text{exc}}$ over the incidence spectrum. Similarly, for a broad emission spectrum from the fluorophore, we should average $\text{ENH}_{\text{coll}}$, $\text{ENH}_{\text{QY}}$, and $\text{ENH}_{1/\tau}$ over the emission spectrum. Unless otherwise stated, we assume a narrow excitation and emission spectrum for simplicity.

Finally, we typically consider as signal enhancement the signal enhancement averaged over axial binding position along the nanowire, that is, $\langle \text{ENH}_{\text{sig}} \rangle = \langle \text{ENH}_{\text{exc}} \text{ENH}_{\text{coll}} \text{ENH}_{\text{QY}} \rangle$ or $\langle \text{ENH}_{\text{sig}} \rangle = \langle \text{ENH}_{1/\tau} \text{ENH}_{\text{coll}} \text{ENH}_{\text{QY}} \rangle$, depending on if we are looking at case (1) or (2) for the excitation. Typically, we can obtain valuable information also from averaging of the constituent terms (Figure S2).

Here, we do not include the bleaching dynamics of fluorophores, in which fluorophores tend to turn into a non-emitting state after a certain number of cycles. However, the above modelling of the modification of the excitation/de-excitation cycle can be used for assessing how much we enhance the bleaching rate in the vicinity of the nanowire, as detailed in our previous work.[4]

## Modelling of fluorescence image

The simulations for the imaging-mode detection are based on the same dipole model as used for the Purcell factor and parasitic absorption above. We use the RETOP package[7] to perform a near-field to far-field transformation (NFFT)—the RETOP package can perform the NFFT even when a substrate is present, as in our case. The far-field emission that enters the collection objective is assumed to be focused onto an image plane. Thus, we assume that the collection objective functions also as the imaging objective such that $\text{NA}_{\text{coll}}$ defines the far-field light that is used in the image formation. Thus, the imaging objective focuses the far-field emission onto the image plane, and the image formation process can be calculated in Fourier space as detailed below.[8]

We describe the far-field emission direction in terms of the $k_x$ and $k_y$ components in the Fourier space, such that $k_x = \sin(\theta)\cos(\varphi) k_0$ and $k_y = \sin(\theta)\sin(\varphi) k_0$ with $\theta$ and $\varphi$ here the polar and azimuth angle that describe the propagation direction (and $k_0 = 2\pi/\lambda$). Then, for each $k_x$-$k_y$ pair, we obtain a phase factor $k_{\text{farfield}}(k_x, k_y)$ that describes the propagation of the far-field component along the optical axis (which is assumed to be parallel to the $z$ axis) and the electric field intensity $\mathbf{E}_{\text{farfield}}(k_x, k_y)$ of the far-field component. The extraction of these far-field components is performed relative to a reference $z$-plane located at $z_{\text{ref}}$. To take into account $z_{\text{focal}}$, the position of the focal plane that we have chosen, we multiply each far-field component with the corresponding shift compared to $z_{\text{ref}}$. That is, we multiply each far-field component with the respective $\exp[ik_{\text{farfield}}(k_x, k_y)(z_{\text{ref}} - z_{\text{focal}})]$.

We use a grid with evenly spaced $k_x$ and $k_y$ in the extraction of far-field direction in RETOP to allow the use of fast-Fourier transform (FFT) to speed up the calculation of the resulting image. In the far-field, we use a cut-off condition given by $\sqrt{(k_x)^2 + (k_y)^2} > \text{NA}_{\text{img}} k_0$, with $\text{NA}_{\text{img}}$ the numerical aperture of the objective used for the imaging (and we assume $\text{NA}_{\text{img}} = \text{NA}_{\text{coll}}$ if not stated otherwise). For such cut-off directions, we set the far-field intensity to zero to still allow the use of FFT. After that, we assume magnification of $M = 1$ for simplicity, which allows us to perform the FFT directly on that far-field spectrum, without additional transformations as would be needed if $M \neq 1$.[8]

Thus, from the FFT we obtain the $E_x$, $E_y$ and $E_z$ that the far-field components give rise to when focused to the image plane. We assume that the corresponding $|\mathbf{E}|^2$ in the image plane yields the signal intensity in the image-based detection. Thus, we implicitly assume that detector in the image plane is polarization independent and with perfect anti-reflection coating, or at least not showing wavelength, incidence angle, or polarization dependent reflection properties.

We denote the result from the above image calculation by $|\mathbf{E}(x,y)_{img,i}|^2$ (where the subscript $i$ denotes that the image originates from dipole with dipole moment oriented in the $i$ direction). Since in our simulations, the far-field component from the NFFT includes modification in the far-field emission intensity due to the modified emission power of the dipole, we normalize with the total power emitted by the dipole—we will instead, explicitly, take into account the modified emission power of the dipole through the modified quantum yield, that is, $\mathrm{ENH}_{QY}$. Furthermore, we take into account the excitation enhancement $\mathrm{ENH}_{exc}$ also similarly as when modelling the signal-integrating detection mode above. For the overall image from the three orthogonal dipole moments, we thus have for the image intensity: $\mathrm{ENH}_{exc}\mathrm{ENH}_{QY}\sum_i \frac{|\mathbf{E}(x,y)_{img,i}|^2}{P_{em,i}} \frac{P_{em,i}}{P_{em,x}+P_{em,y}+P_{em,z}}$ where the second fraction gives the probability that the emission occurs through dipole orientation $i$. The above scheme for the imaging modelling works for the various systems included in our current study, i.e., for the nanowire-system, the planar substrate-system without nanowire, and the fluorophore in the homogeneous liquid surrounding.

## Electromagnetic simulations

We perform the solving of the Maxwell equations with the finite element method in Comsol Multiphysics for the three-dimensional (3D) geometry of the vertical nanowire on top of a substrate, covered by an oxide layer. We consider two types of modelling: (i) the scattering of an incident plane wave to yield the $\mathbf{E}(\mathbf{r}, \lambda, \theta_{inc}, \phi_{inc}, pol)$ for Eqs. (S3) and (S4) and (ii) emission from a dipole emitter as needed for the Purcell factor and analysis of parasitic absorption. From the dipole emitter simulation, we also obtain with the RETOP package[7] the NFFT, which is needed for the image creation when analysing detection in imaging mode.

In all the simulations, for simplicity, we set the imaginary part of the refractive index of the substrate to zero. This simplification has only a very minor effect on the results as seen in Figure S1.

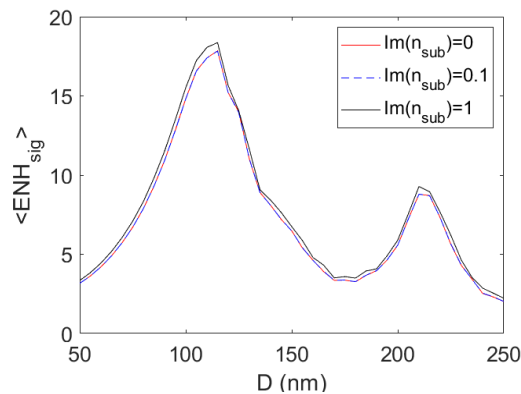

**Figure S1.** Signal enhancement for a nanowire with $n$ = 3.5 when $\lambda_{exc}$ = 640 nm, $\lambda_{em}$ = 670 nm, $QY_0$ = 0.33, and $NA_{exc}$ = $NA_{coll}$ = 1. Here, the substrate has refractive index with $\mathrm{Re}(n_{sub})$ = 3.5 and varying $\mathrm{Im}(n_{sub})$. Even at $\mathrm{Im}(n_{sub})$ = 1 for the substrate, the difference is rather minor as compared to the results with $\mathrm{Im}(n_{sub})$ = 0. There is a slight increase in the signal with increasing $\mathrm{Im}(n_{sub})$, which can be understood from the expectation of an increased reflection of both excitation light and emitted light at the substrate interface with increasing refractive index (at least for light propagating at close to normal incidence, as given by the Fresnel equations for a planar interface). To compare this behaviour for increasing $\mathrm{Im}(n_{sub})$ for the substrate with the refractive index values of the materials that we consider in this study and that absorb in the wavelength range studied: As seen in Figure S3, $\mathrm{Im}(n_{sub})$ < 0.1 for $\lambda$ > 450 nm and $\mathrm{Im}(n_{sub})$ < 0.3 for 400 < $\lambda$ < 450 nm for Si and GaAs. For GaAs, $\mathrm{Im}(n_{sub})$ < 1 for $\lambda$ > 445 nm and $\mathrm{Im}(n_{sub})$ < 2.15 for 400 nm < $\lambda$ < 445 nm.

## Effect of averaging order

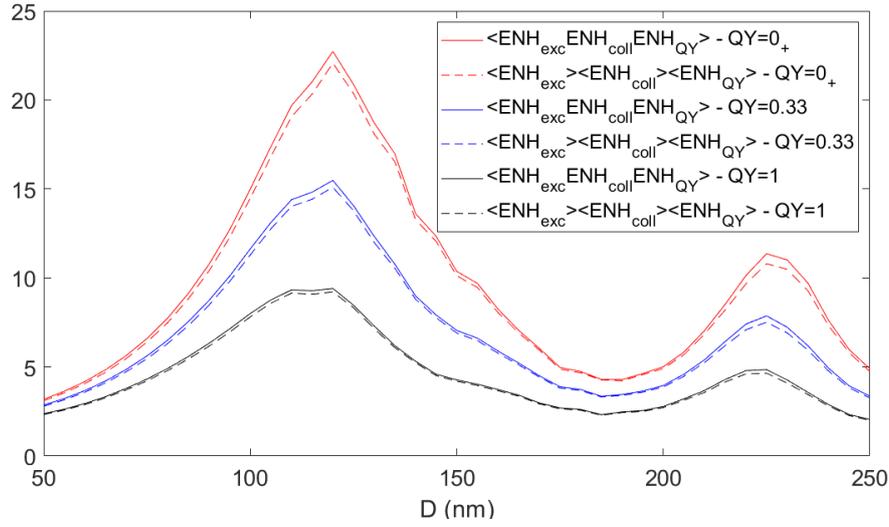

**Figure S2**. Signal enhancement for fluorophores at the side wall of a GaP nanowire of $L$ = 2000 nm in length with a 10 nm thick SiO$_2$ coating, averaged over axial position, that is, $\langle \text{ENH}_{\text{sig}}\rangle = \langle \text{ENH}_{\text{exc}}\text{ENH}_{\text{coll}}\text{ENH}_{\text{QY}}\rangle$ compared to the case where each of the three components is first averaged over axial position and the resulting averages are multiplied together, that is, $\langle \text{ENH}_{\text{exc}}\rangle\langle \text{ENH}_{\text{coll}}\rangle\langle \text{ENH}_{\text{QY}}\rangle$. These two ways of averaging give rather similar results, indicating that we can explain the signal enhancement process accurately by looking at the axially-averaged value of each of the three contributing factors separately. Here, $\lambda_{\text{exc}}$ = 640 nm, $\lambda_{\text{em}}$ = 670 nm, QY$_0$ = 0.33, and NA$_{\text{exc}}$ = NA$_{\text{coll}}$ = 1

## Refractive index for the modelled materials

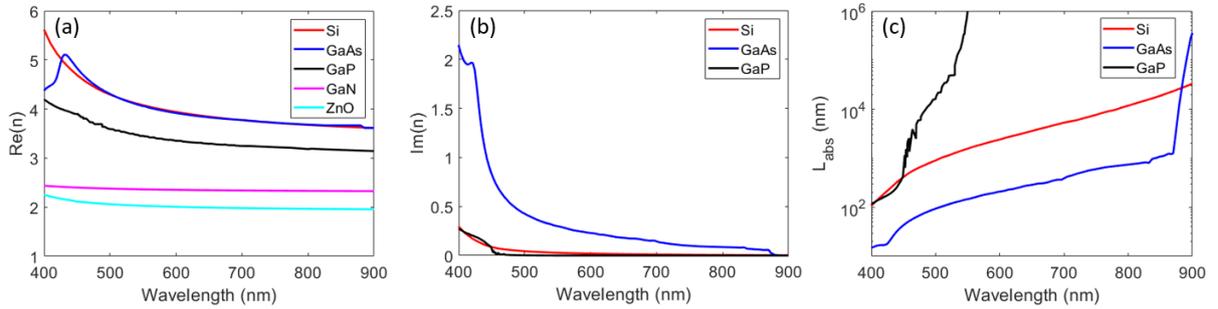

**Figure S3.** (a) Real part and (b) imaginary part of the refractive index of the nanowire materials considered in our study. (c) Corresponding absorption length $L_{\text{abs}} = (4\pi\text{Im}(n)/\lambda)^{-1}$ in the materials—GaN and ZnO do not show absorption in this wavelength range, that is, $\text{Im}(n) = 0$ for them.

We consider the wavelength range from 400 to 900 nm, which is a broad wavelength range covering many fluorophores of practical interest.[9] For the water on top of and around the nanowire, we assume a fixed refractive index of $n$ = 1.33. For the oxide coating, we use fused silica, for which the refractive index drops from 1.47 at $\lambda = 400$ nm to 1.45 at $\lambda = 900$ nm .[10]

We show in Figure S3 the tabulated refractive index values for Si,[11] GaAs,[12] GaP,[13] GaN,[14] and ZnO[15] used in the modelling—for ZnO, we extend the range to wavelengths below 450 nm with values[16] from a model. Both ZnO and GaN show birefringence in their refractive index, with a slight difference between the ordinary and extraordinary refractive index, $n_{\text{o}}$ and $n_{\text{e}}$. For ZnO, the difference between $n_{\text{o}}$ and $n_{\text{e}}$ is less than 0.02.[15] For GaN, the difference is 0.25 at $\lambda = 400$ nm, decreasing monotonically to 0.1 at $\lambda = 550$ nm and 0.03 at $\lambda = 900$ nm.[14] For simplicity, we use in the simulations for the refractive index of ZnO and GaN the average of their $n_{\text{o}}$ and $n_{\text{e}}$, and the resulting values are shown in Figure S3(a).

# Signal enhancement away from nanowire sidewall

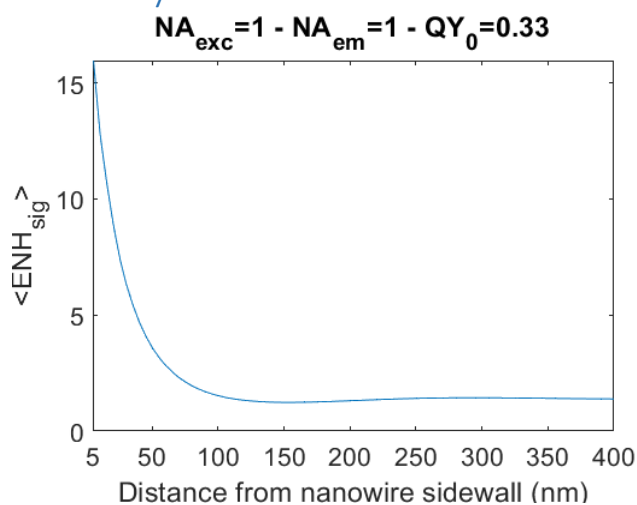

**Figure S4**. Signal enhancement from Figure 2(a), here averaged over *z*-position for the length of the nanowire, given in terms of the distance from the nanowire sidewall, with a value of zero corresponding to the interface between the oxide coating and the water. For large distance, we find a value of $\langle \mathrm{ENH}_{\mathrm{sig}} \rangle \approx 1.4$ where the difference to unity (that applies for homogeneous liquid) is due to the presence of the substrate that causes reflection of light.

# Collection efficiency

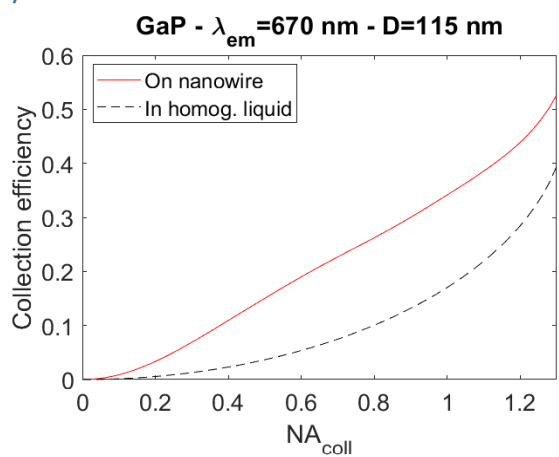

**Figure S5**. Collection efficiency for $NA_{\mathrm{coll}} < 1.3$, that is, the fraction of emitted photons collected into $NA_{\mathrm{coll}}$. Here, averaging over axial position on the nanowire surface is performed. For the nanowires, $D$ = 115 nm and $L$ = 2000 nm is used. We show also the corresponding value for the fluorophore in the homogeneous test liquid (dashed line) where the value goes toward 0.5 when $NA_{\mathrm{coll}} \to 1.33$, in which case the collection is from the whole top hemisphere since $n$ = 1.33 is used for the refractive index of the liquid.

## Detuning between excitation and emission wavelength

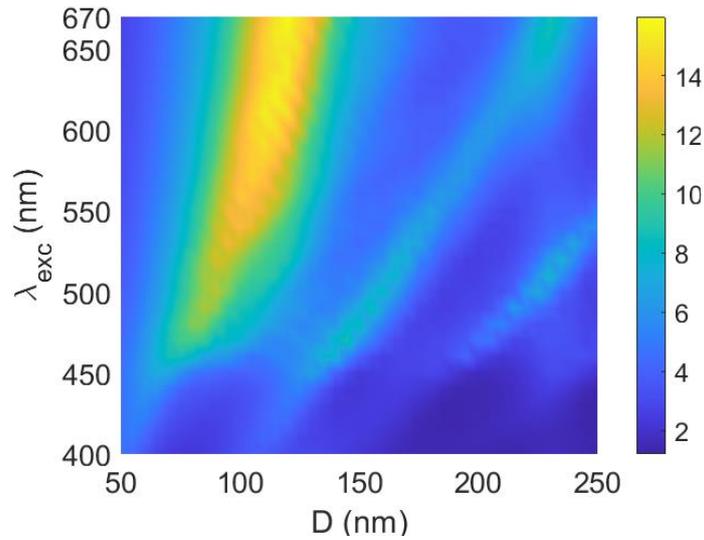

**Figure S6**. Signal enhancement from fluorophores, averaged over axial position, that is, $\langle \text{ENH}_{\text{sig}} \rangle = \langle \text{ENH}_{\text{exc}} \text{ENH}_{\text{coll}} \text{ENH}_{\text{QY}} \rangle$, as a function of nanowire diameter and excitation wavelength. Here, $QY_0$ = 0.33, $NA_{\text{exc}} = 1$, $NA_{\text{coll}} = 1$, and $\lambda_{\text{em}} = 670$ nm. The nanowires are of 2000 nm in length and of GaP, the substrate is of GaP, and there is a 10 nm thick SiO$_2$ coating on the substrate and nanowire surface.

## Signal enhancement for varying oxide coating thickness

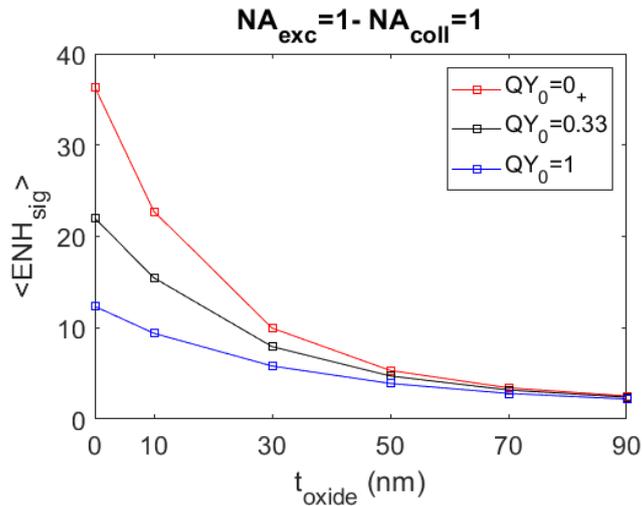

**Figure S7.** Signal enhancement, $\langle \text{ENH}_{\text{sig}} \rangle$, for optimum nanowire diameter, as a function of thickness, $t_{\text{oxide}}$, of the SiO$_2$ coating. $NA_{\text{exc}} = 1$, $NA_{\text{coll}} = 1$, $\lambda_{\text{exc}} = 640$ nm, and $\lambda_{\text{em}} = 670$ nm. For $t_{\text{oxide}} = 0$, the optimum diameter is 125 nm for all considered $QY_0$. With increasing $t_{\text{oxide}}$, the optimum diameter decreases, and at $t_{\text{oxide}} = 90$ nm, it is 90 nm for $QY_0 = 0_+$ and 0.33 and 85 nm for $QY_0 = 1$. The nanowires are of 2000 nm in length and of GaP. The substrate is of GaP.

## Signal enhancement for varying nanowire length

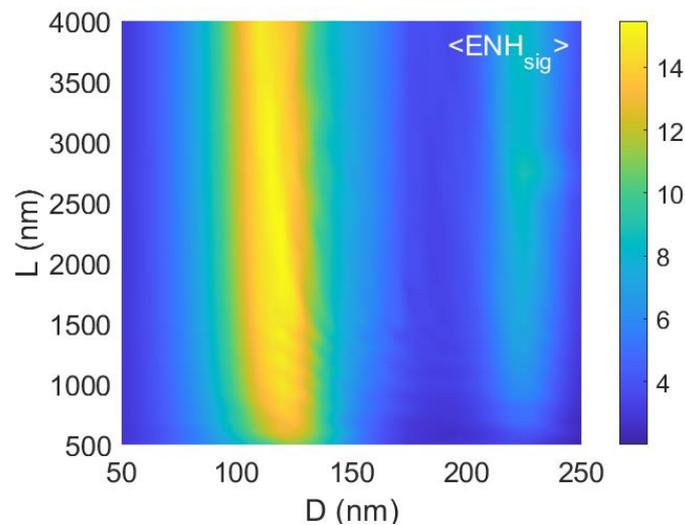

**Figure S8.** Signal enhancement as a function of nanowire diameter and length. Here, $QY_0 = 0.33$, $NA_{exc} = 1$, $NA_{coll} = 1$, $\lambda_{exc} = 640$ nm, and $\lambda_{em} = 670$ nm. The nanowires are of GaP, the substrate is of GaP, and there is a 10 nm thick SiO$_2$ coating on the substrate and nanowire surface.

## Signal enhancement for varying Re(n) and Im(n) for nanowire material

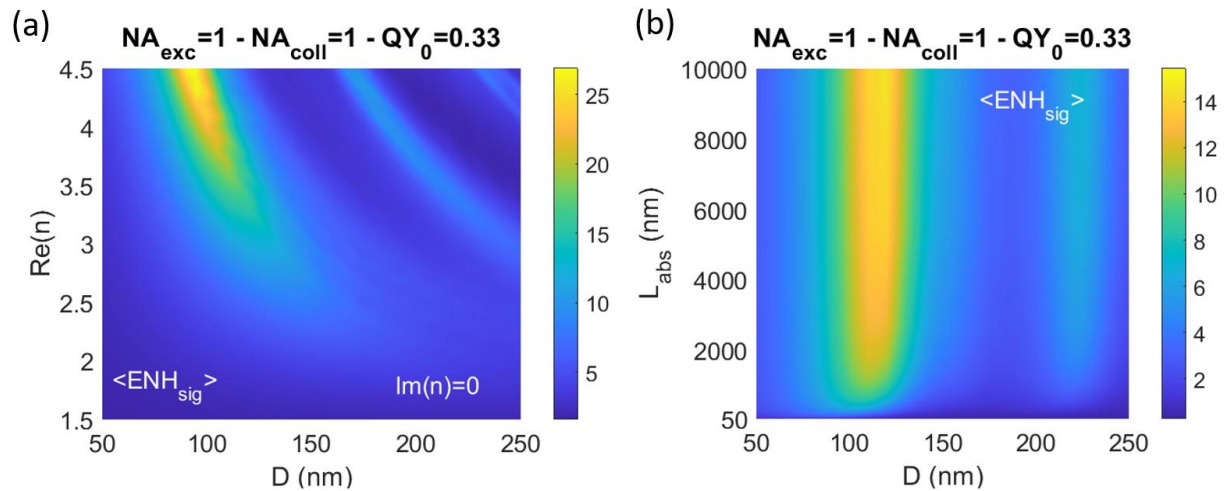

**Figure S9.** (a) Signal enhancement as a function of nanowire diameter and (wavelength independent) Re(n) of nanowire material when Im(n) = 0 is set for the nanowire material. (b) Signal enhancement as a function of nanowire diameter and (wavelength independent) Im(n) of nanowire material, here expressed in terms of absorption length $L_{abs} = (4\pi \text{Im}(n)/\lambda)^{-1}$ at the wavelength of 670 nm. In (b), we use Re(n) = 3.302 at $\lambda = \lambda_{exc}$ and Re(n) = 3.272 at $\lambda = \lambda_{em}$, corresponding to the values of Re(n) for GaP. In both (a) and (b), $QY_0 = 0.33$, $NA_{exc} = 1$, $NA_{coll} = 1$, $\lambda_{exc} = 640$ nm, $\lambda_{em} = 670$ nm, the nanowires are of 2000 nm in length, the substrate is of GaP, and there is a 10 nm thick SiO2 coating on the substrate and nanowire surface.

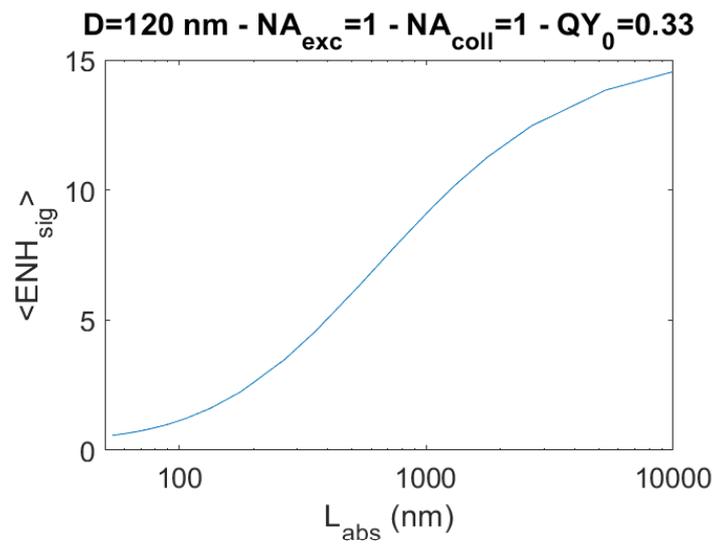

**Figure S10.** Line-cut from Figure S9(b) at *D* = 120 nm.

# Signal enhancement for GaP, Si, GaAs, ZnO, and GaN nanowires

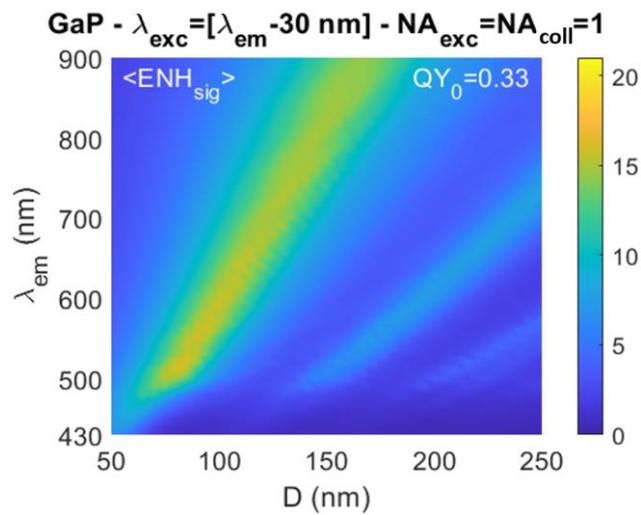

**Figure S11**. As Figure 6b, but here the colour scale is set to match that in Figure S12.

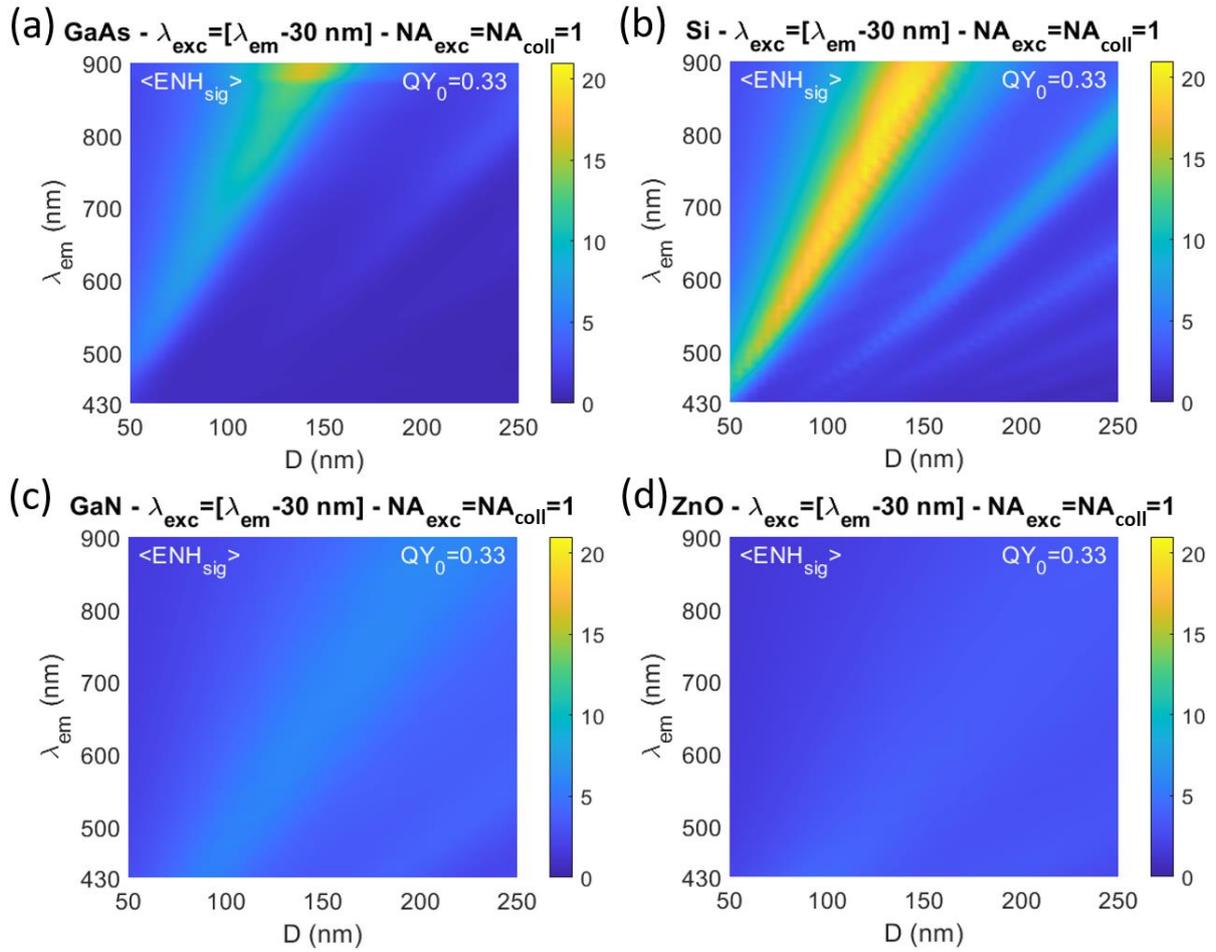

**Figure S12**. Signal enhancement as a function of emission wavelength and nanowire diameter from fluorophores, averaged over axial position, that is, $\langle \text{ENH}_{sig} \rangle$, for (a) GaAs, (b) Si, (c) GaN, and (d) ZnO nanowires of $L$ = 2000 nm in length with a 10 nm thick SiO$_2$ coating. The nanowire is placed on a substrate of the same material as the nanowire. Here, $\text{NA}_{exc} = \text{NA}_{em} = 1$, the excitation wavelength is related to the emission wavelength through a 30 nm offset, that is, $\lambda_{exc} = \lambda_{em} - 30$ nm, and we assume $\text{QY}_0 = 0.33$. Here, $\langle \text{ENH}_{sig} \rangle$ for all the materials is shown on the same color scale for easy comparison between materials. For a view of GaN and ZnO on a rescaled colour axis that easier shows the values, see Figure S13.

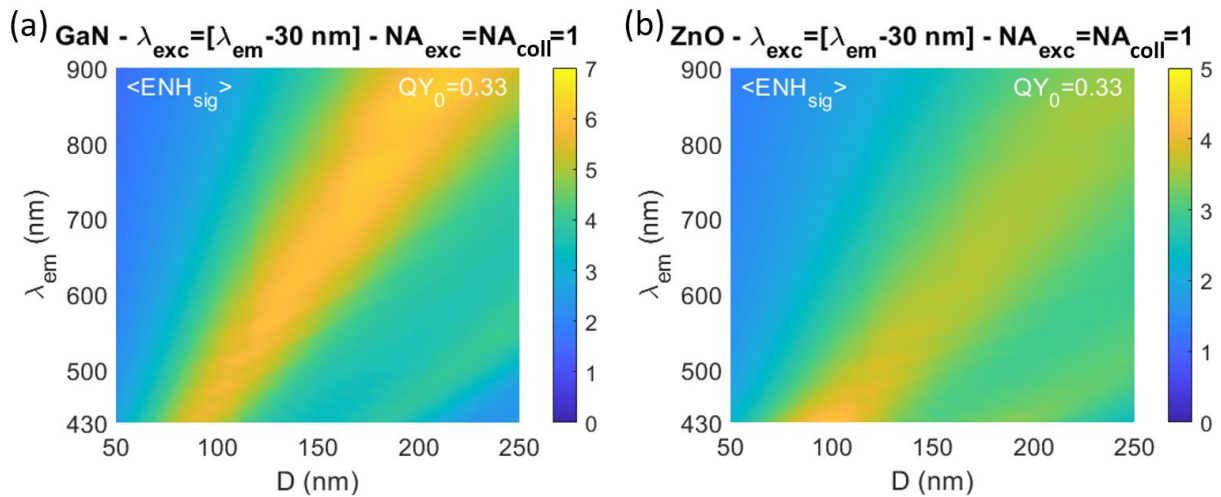

**Figure S13**. Same as Figure S12(c,d) but with rescaled colour scale.

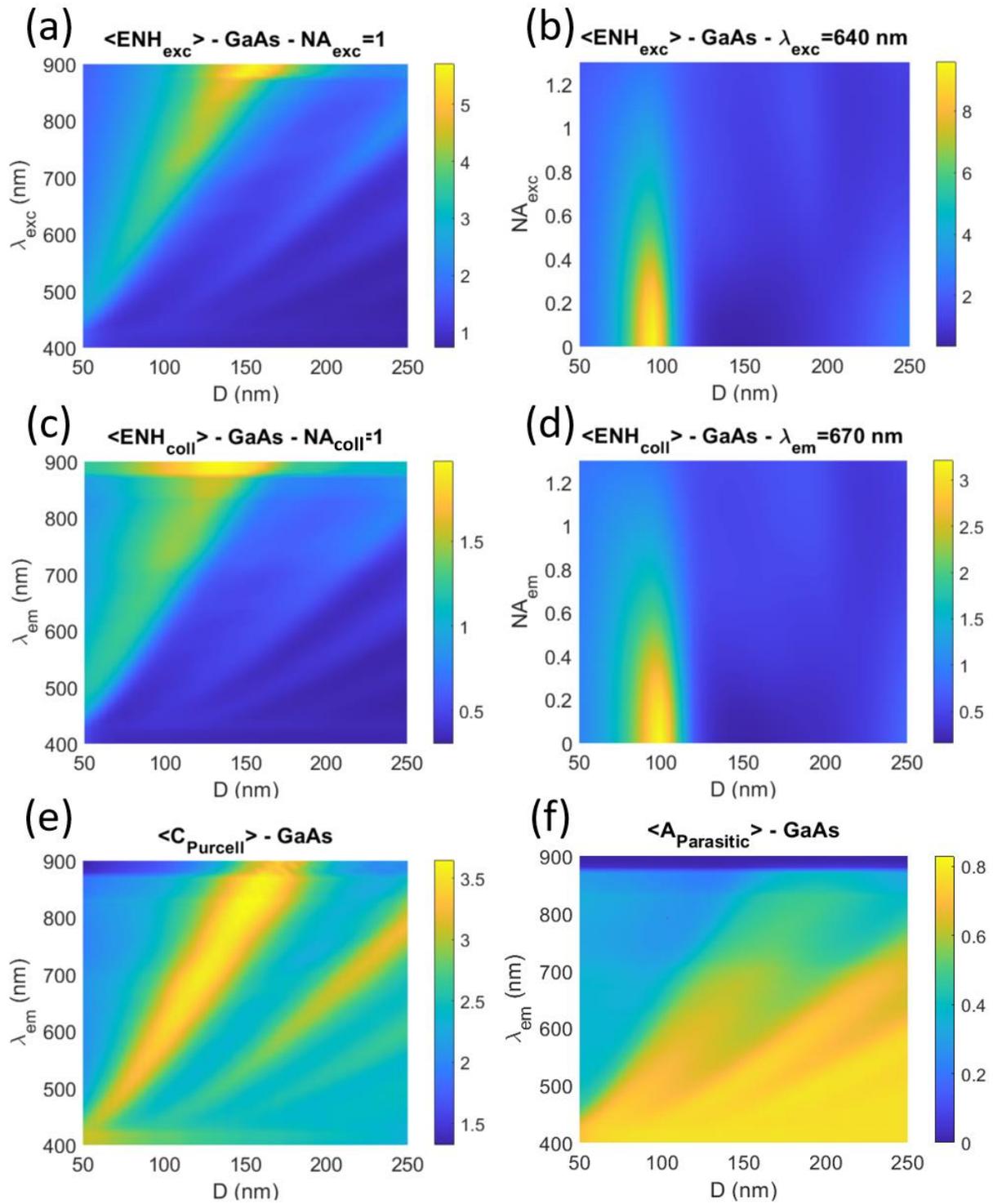

**Figure S14.** (a)-(b) Enhancement of incident intensity for fluorophores at the side wall of a GaAs nanowire of $L$ = 2000 nm in length with a 10 nm thick $SiO_2$ coating. (c)-(d) Enhancement of the collection of light from the fluorophore. (e) Modification of the Purcell factor. (f) Parasitic absorption in the nanowire of emitted light from the fluorophore.

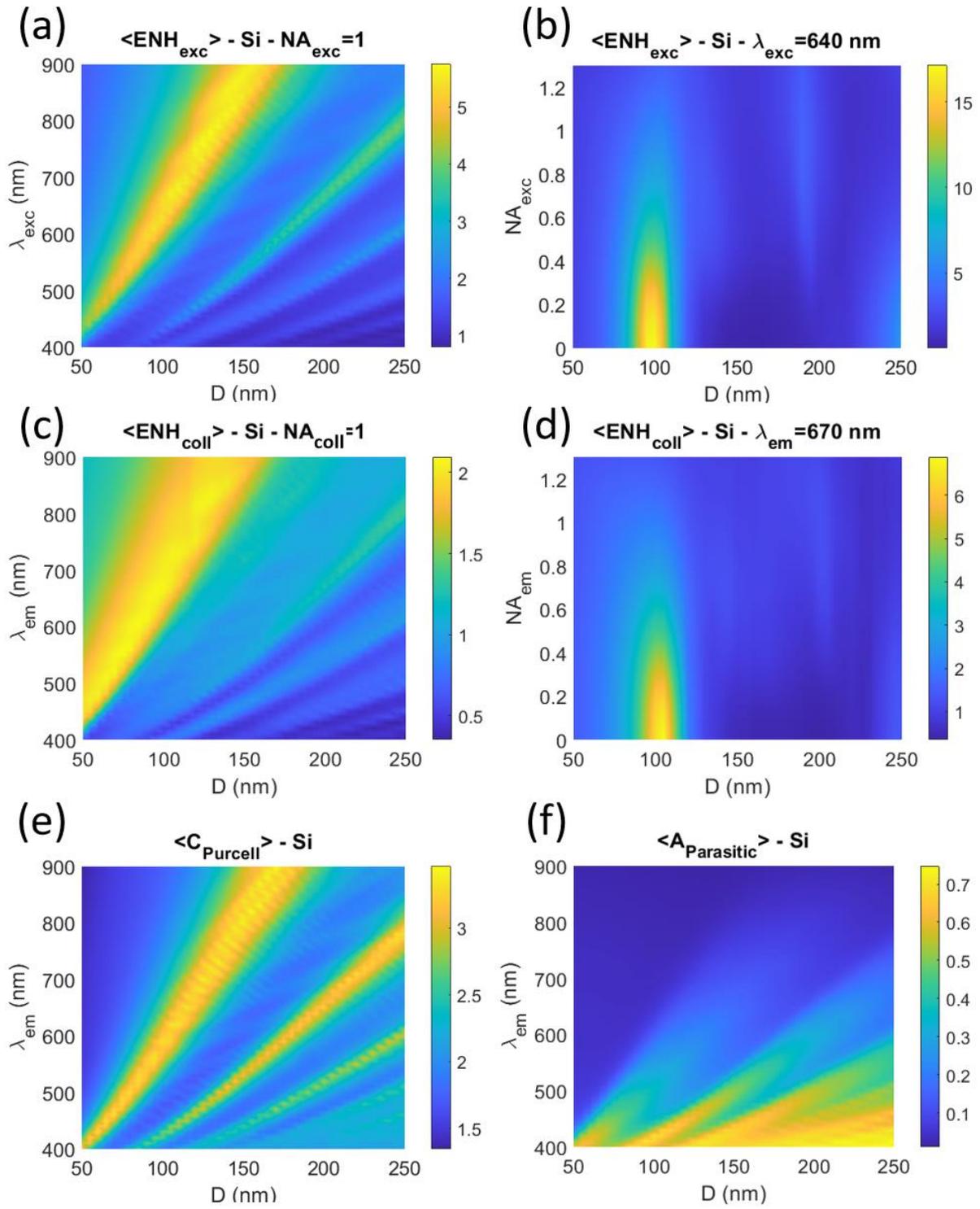

**Figure S15**. (a)-(b) Enhancement of incident intensity for fluorophores at the side wall of a Si nanowire of $L$ = 2000 nm in length with a 10 nm thick $SiO_2$ coating. (c)-(d) Enhancement of the collection of light from the fluorophore. (e) Modification of the Purcell factor. (f) Parasitic absorption in the nanowire of emitted light from the fluorophore.

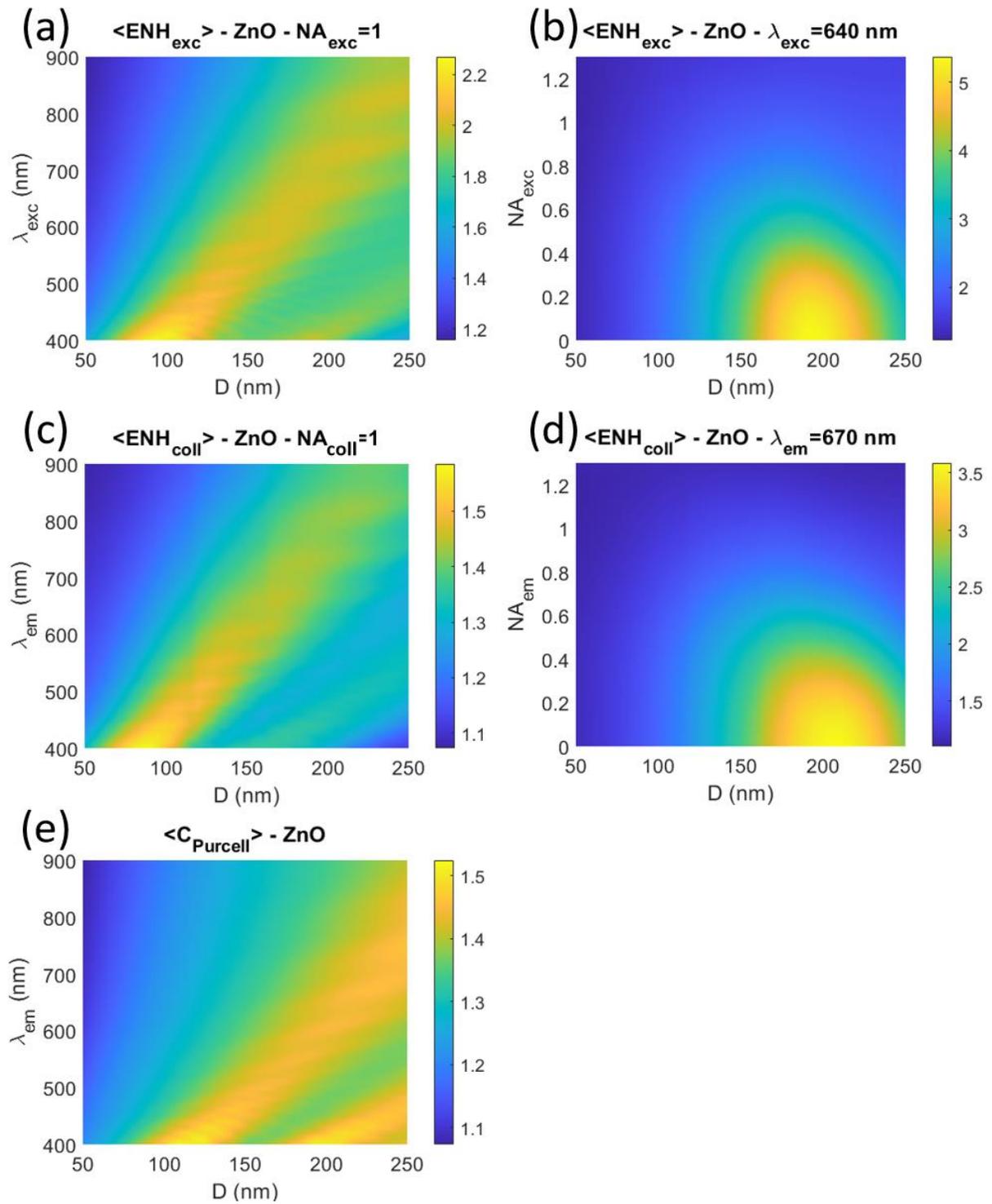

**Figure S16**. (a)-(b) Enhancement of incident intensity for fluorophores at the side wall of a ZnO nanowire of $L$ = 2000 nm in length with a 10 nm thick $SiO_2$ coating. (c)-(d) Enhancement of the collection of light from the fluorophore. (e) Modification of the Purcell factor. Note that in this wavelength range, Im($n_{ZnO}$) = 0, and therefore parasitic absorption in the nanowire is zero.

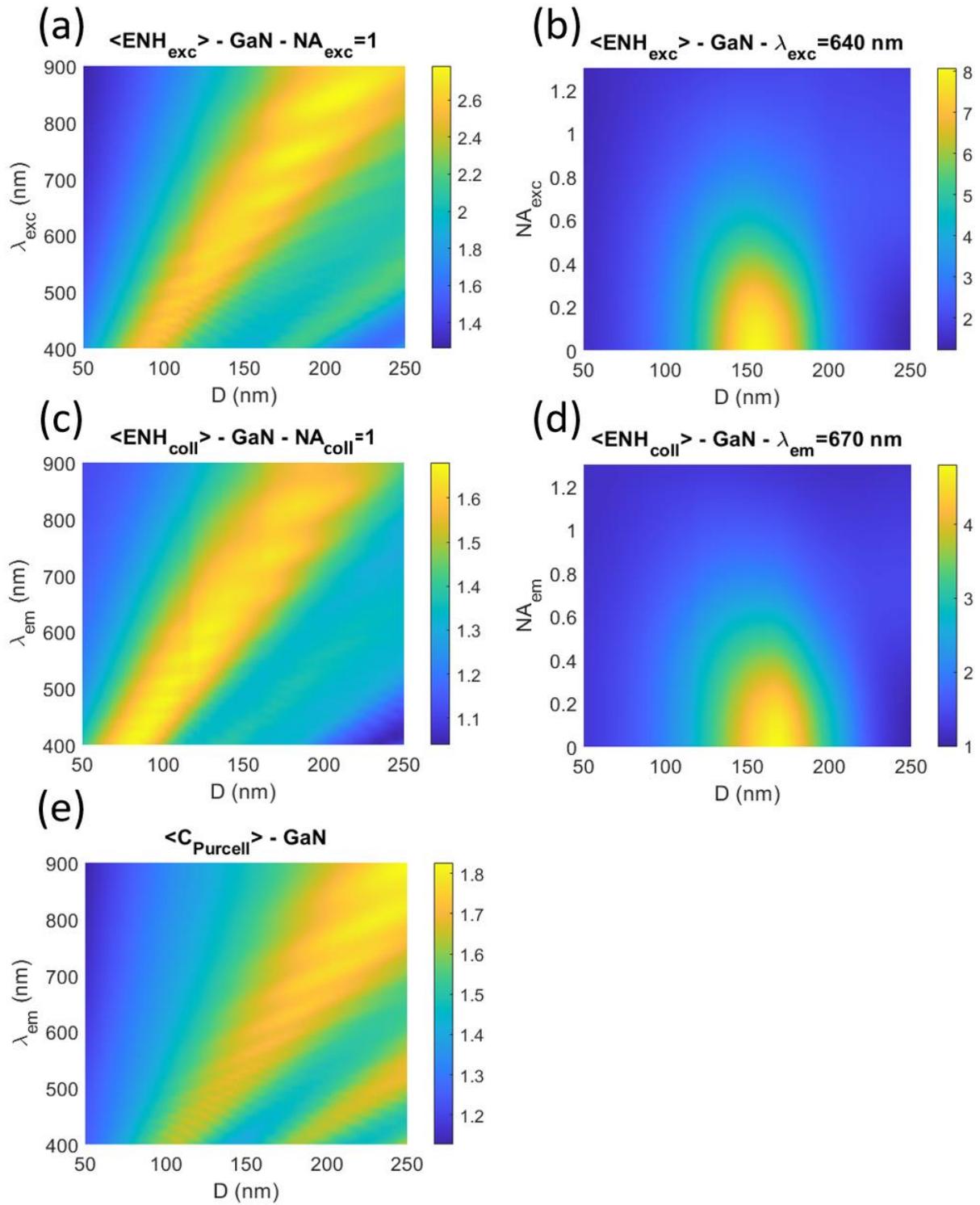

**Figure S17.** (a)-(b) Enhancement of incident intensity for fluorophores at the side wall of a GaN nanowire of $L$ = 2000 nm in length with a 10 nm thick $SiO_2$ coating. (c)-(d) Enhancement of the collection of light from the fluorophore. (e) Modification of the Purcell factor. Note that in this wavelength range, Im($n_{GaN}$) = 0, and therefore parasitic absorption in the nanowire is zero.

## Enhancement in imaging mode

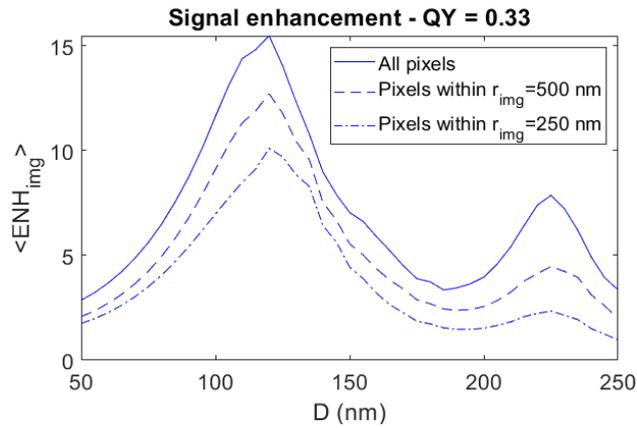

**Figure S18.** Signal enhancement in imaging mode, averaged over axial binding position for the fluorophore, that is, $\langle \text{ENH}_{\text{img}} \rangle$. We consider two cases: (1) Integration of signal over all pixels in the image plane (solid line), which is equivalent to the non-imaging detection case $\langle \text{ENH}_{\text{sig}} \rangle$ that detects (or at least is proportional to) all photons within $\text{NA}_{\text{coll}}$; and (2) Integration over pixels within $r_{\text{img}} = 500$ nm from the centre of the nanowire in the image (dashed line), as well as within $r_{\text{img}} = 250$ nm (dashed-dotted line). Here, $QY_0 = 0.33$, $\text{NA}_{\text{exc}} = 1$, $\text{NA}_{\text{coll}} = 1$, $\lambda_{\text{exc}} = 640$ nm, and $\lambda_{\text{em}} = 670$ nm. The nanowire is of 2000 nm in length and of GaP, the substrate is of GaP, and there is a 10 nm thick $SiO_2$ coating on the substrate and nanowire surface. For the cases of the detection disc of $r_{\text{img}} = 250$ and 500 nm, for each value of $D$, the focal plane is placed at optimum position—see Figure 7(b) for the optimum $z_{\text{focal}}$ for $r_{\text{img}} = 500$ nm—that is, the values shown here for $r_{\text{img}} = 500$ nm with the dashed line are the maximum values from Figure 7(a) for a given $D$.

## Defocusing in non-nanowire systems

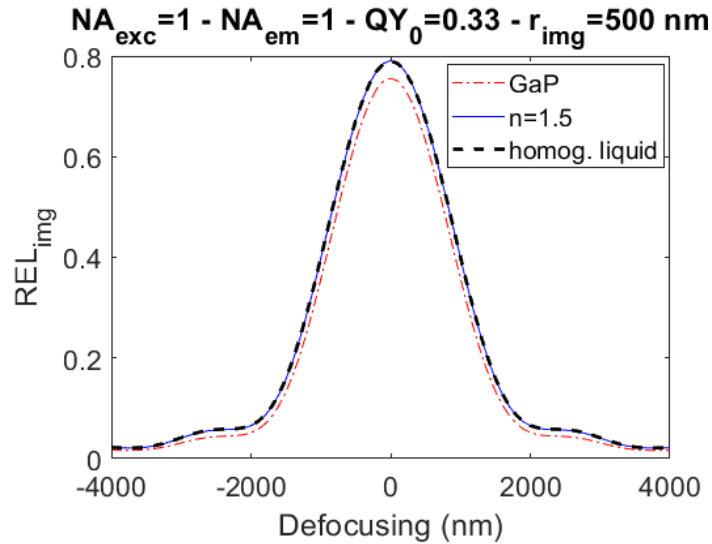

**Figure S19.** Integrated counts in a disc of 500 nm in radius in the image plane, relative to all counts in the image plane, that is, $\text{REL}_{\text{img}}$ for $r_{\text{img}} = 500$ nm. We consider here a planar GaP substrate system with the fluorophore on top of the substrate (red dashed-dotted line), a $n = 1.5$ substrate system (blue solid line) with the fluorophore on top of the substrate, and a homogeneous $n = 1.33$ surrounding (dashed line). Positive defocusing corresponds, for the substrate systems, to a shift to above the substrate surface. Here, $QY_0 = 0.33$, $\text{NA}_{\text{exc}} = 1$, $\text{NA}_{\text{coll}} = 1$, $\lambda_{\text{exc}} = 640$ nm, and $\lambda_{\text{em}} = 670$ nm. As reference value for the overall signal, the signal enhancement in non-imaging mode for the GaP substrate system is 0.24, while it is 0.78 for the $n = 1.5$ substrate system.

## Varying nanowire length in imaging detection mode

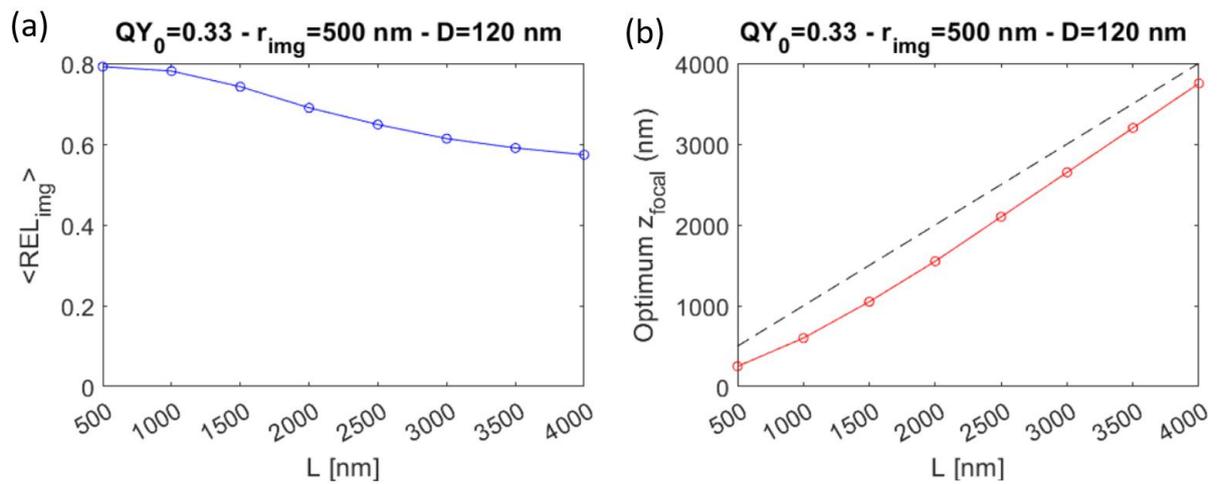

**Figure S20.** (a) Integrated counts in a disc of 500 nm in radius in the image plane, centred on the nanowire, relative to all counts in the image plane, averaged over fluorophore binding position along the nanowire axis, that is, $\langle \mathrm{REL}_{\mathrm{img}} \rangle$ for $r_{\mathrm{img}} = 500$ nm. The results in (a) are shown for optimum $z_{\mathrm{focal}}$ for each $L$. The corresponding optimum $z_{\mathrm{focal}}$ (red circles) is shown in (b), with the dashed line showing $z_{\mathrm{focal}} = L$. Here, $\mathrm{NA}_{\mathrm{exc}} = \mathrm{NA}_{\mathrm{coll}} = 1$, $\lambda_{\mathrm{exc}} = 640$ nm, $\lambda_{\mathrm{em}} = 670$ nm, and $\mathrm{QY}_0 = 0.33$. The nanowires are of GaP and $D = 120$ nm. The substrate is of GaP, and there is a 10 nm thick $SiO_2$ coating on the substrate and nanowire surface.

# Lightguiding and non-lightguiding component in image creation

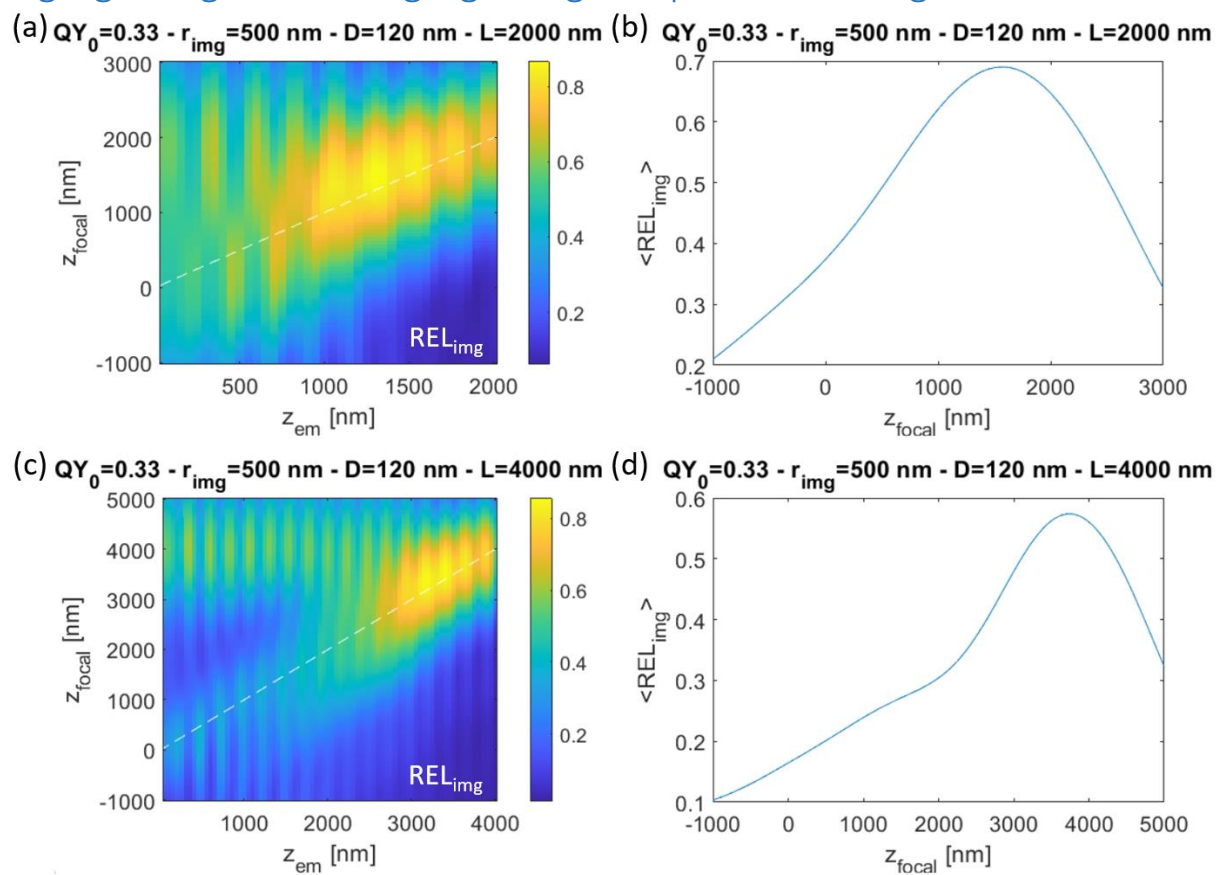

**Figure S21.** Integrated counts in a disc of 500 nm in radius in the image plane, relative to all counts in the image plane. (a) and (c) show results as a function of $z_{em}$, the position of the fluorophore on the sidewall of the nanowire, and $z_{focal}$, the focal plane for the imaging, that is, (a) and (c) show $\text{REL}_{img}$ for $r_{img} = 500$ nm. (b) and (d) show values averaged over $z_{em}$ from (a) and (c), respectively, that is, (b) and (d) show $\langle \text{REL}_{img} \rangle$ for $r_{img} = 500$ nm. Here, $\text{NA}_{exc} = \text{NA}_{coll} = 1$, $\lambda_{exc} = 640$ nm, $\lambda_{em} = 670$ nm, and $QY_0 = 0.33$. The nanowires are of GaP and $D = 120$ nm. In (a) and (b), $L = 2000$ nm, and in (c) and (d), $L = 4000$ nm. The substrate is of GaP, and there is a 10 nm thick $SiO_2$ coating on the substrate and nanowire surface. The straight dashed line in (a) and (c) corresponds to focal plane at the position of the fluorophore, that is, $z_{focal} = z_{em}$. The contribution along this line indicates non-guided light, for which it is beneficial to focus to the actual position of the fluorophore. In contrast, the peak in the contribution at $z_{focal} = L$, which is especially visible at the larger $L = 4000$ nm in (c), indicates light-guiding where light appears for imaging as emitted at the top of the nanowire, irrespective of the actual binding position of the fluorophore.

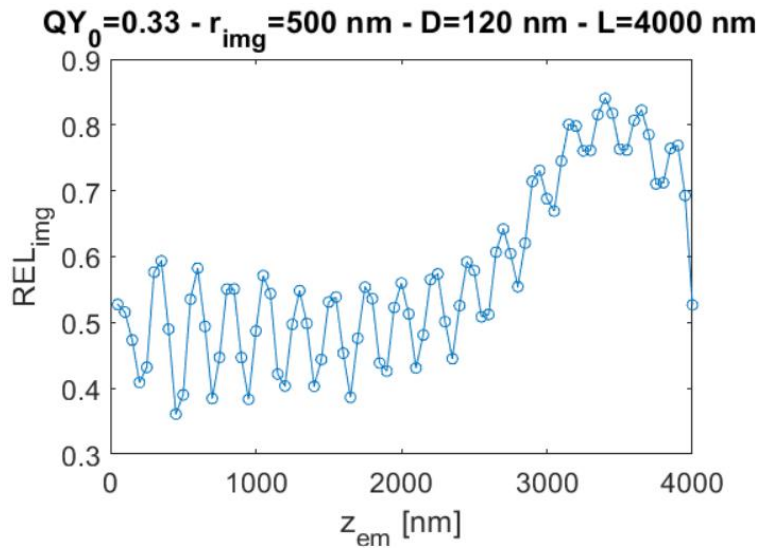

**Figure S22.** Results extracted from Figure S21(c) at the optimum $z_{\text{focal}}$ = 3750 nm.